# Technical Report: NUS-ACT-11-005-Ver.1
# Hybrid 3-D Formation Control for Unmanned Helicopters☆

Ali Karimoddini, Hai Lin, Ben. M. Chen, Tong Heng Lee

*National University of Singapore*

**Abstract**

Teams of Unmanned Aerial Vehicles (UAVs) form typical networked cyber-physical systems that involve the interaction of discrete logic and continuous dynamics. This paper presents a hybrid supervisory control framework for the three-dimensional leader follower formation control of unmanned helicopters. The proposed hybrid control framework captures internal interactions between the decision making unit and the path planner continuous dynamics of the system, and hence improves the system's overall reliability. To design such a hybrid controller, a spherical abstraction of the state space is proposed as a new method of abstraction. Utilizing the properties of multi-affine functions over the partitioned space leads to a finite state Discrete Event System (DES) model, which is shown to be bisimilar to the original continuous-variable dynamical system. Then, in the discrete domain, a logic supervisor is modularly designed for the abstracted model. Due to the bisimilarity between the abstracted DES model and the original UAV dynamics, the designed logic supervisor can be implemented as a hybrid controller through an interface layer. This supervisor drives the UAV dynamics to satisfy the design requirements. In other words, the hybrid controller is able to bring the UAVs to the desired formation starting from any initial state inside

☆The authors are with National University of Singapore, Graduate School for Integrative Sciences and Engineering (NGS), Centre for Life Sciences (CeLS), #05-01, 28 Medical Drive, 117456 Singapore and the Department of Electrical and Computer Engineering, National University of Singapore, Engineering Drive 3, 117576 Singapore.
Corresponding author: H. Lin (elelh@nus.edu.sg), Tel. (+65) 6516-2575 . Fax (+65) 6779-1103.

the control horizon and then, maintain the formation. Moreover, a collision avoidance mechanism is embedded in the designed supervisor. Finally, the algorithm has been verified by a hardware-in-the-loop simulation platform, which is developed for unmanned helicopters. The presented results show the effectiveness of the algorithm.

# 1. Introduction

Formation control of the Unmanned Aerial Vehicles (UAVs) is a typical cooperative control scenario in which a group of UAVs jointly moves with fixed relative distances ([1], [2], [3]). Formation control can leverage limited capabilities of individual UAVs so that as a team, they are able to perform complicated tasks such as mapping, search, coverage, surveillance, and transportation, cooperatively. They can also mutually support each other in a hostile or hazardous environment ([1], [2], [3]).

In general, a formation scenario consists of several subtasks. Firstly, the UAVs should be controlled to form the desired formation. After reaching the formation, the UAVs should maintain the achieved formation while as a team, they have to track a desired path. Moreover, to guarantee the mission safety, the control algorithm is required to take care of inter-collision between the vehicles.

To address these problems, different methods have been developed in the literature. For reaching the formation, there are several existing methods such as optimal control techniques, navigation function, and potential field ([4], [5], [6] , [7]). Maintaining the formation can be seen as a standard control problem in which the system's actual position has slightly deviated from the desired position ([8], [9], [10], [11], [12]). Finally, in [13], [14], [15], [16], and [17], different mechanisms for collision avoidance have been introduced using predictive control, probabilistic methods, MILP programming, invariant sets, and behavioral control.

To obtain a complete solution for the formation control, one way is to separately design a controller for each of these tasks in a decoupled way, and then, a decision making unit is needed to orchestrate the switching between these subcontrollers according to different scenarios. Although the negligence of the interaction between the continuous dynamics and the discrete logic de-cisions simplifies the design, this may degrade the reliability of the overall system, and may cause unexpected failures as it happened in Arian 5 [18]. Hence, a more comprehensive analysis requires an in-depth understanding of the interplay between the continuous dynamics and the discrete supervisory logic of the system. One suitable solution for this problem is to utilize the hybrid modelling and control theory ([19], [20], [21]) that provides a mathematic framework to capture both the continuous and the discrete dynamics of the system, simultaneously.

In this paper, we propose a 3-D hybrid supervisory control architecture for the path planner layer of the UAV helicopters that are involved in a leader-follower formation mission. Furthermore, a new method of abstraction is introduced, which uses the spherical partitioning of the state space. Utilizing multi-affine functions over the partitioned space, the designer is able to make each partition as an invariant set or to deterministically drive the UAV towards one of its adjacent partitions. The proposed abstracted model of the system can be captured by a finite state machine, which can be further studied through the Discrete Event Systems (DES) supervisory control theory initiated by Ramadge and Wonhom [22]. Within the DES framework, we can modularly design the discrete supervisors for reaching the formation, keeping the formation, and avoiding the collision.

To implement the designed discrete supervisor, inspiring from [19], an interface layer has been constructed to link the discrete supervisor and the continuous plant. It can be shown that the discrete model of the plant and its partitioned model have a bisimulation relation, so that they can exhibit the same behavior.

The proposed method is based on spherical partitioning of the state space using the properties of multi-affine functions. Indeed, multi-affine functions are a large class of systems that are decidable under triangulization or rectangulization of the state space ([23], [24], [25]). So far, these methods have not been used in the UAV path planning and formation control. Moreover, the direct path towards the desired point is not available in a reactangulized or trangulized space. In [26], we proposed a 2-D formation algorithm based on polar abstraction of the state space for which the direct path towards the target is applicable, which facilitates the implementation and the design of the formation algorithm. However, in [26], it had been assumed that the UAV's altitude remain unchanged and irrelevant. Compared with [26], the main contributions in this paper are that firstly, the results are extended to the 3-D space and the spherical partitioning of the state space is provided. The extension of polar abstraction to the spherical abstraction is not trivial as the structure of sectors in polar and spherical coordinate systems are different. Moreover, the resulting DES models are different. Secondly, to show that the designed controller for the abstract model works for the original plant, the bisimulation of the partitioned model of the plant and its abstracted DES model is proved and the bisimulation relation is accurately described. Thirdly, the algorithm has been verified through the hardware-in-the-loop simulation platform developed by our research group.

The rest of this paper is organized as follows. First, the problem of formation control is formulated in Section 2, and then, Section 3 describes the principles of the spherical partitioning of the state space. Section 4 utilizes the properties of multi-affine functions over the partitioned state space. In Section 5, two important control features, the invariant region and the exit facet, are introduced. Section 6 abstracts the partitioned model into a finite state machine and proves the bisimulation relation between the partitioned model of the plant and its abstract model. In Section 7, the DES model of the system has been developed and a supervisor has been modularly designed. to handle reaching the formation, keeping the formation, and collision avoidance, accordingly. Section 8 describes how the discrete supervisor can be applied to the plant and how the closed-loop system works. Section 9 demonstrates the hardware-in-the-loop simulation results to verify the proposed algorithm. The paper is concluded in Section 10.

## 2. Problem formulation

The modelling and low-level control structure of the NUS UAV helicopter is explained in [27] and [28]. This UAV is a Raptor-90 helicopter with 1410 mm the full length and 190 mm full width of fuselage [29]. For this helicopter we have used a hierarchical control structure whose inner-loop controller stabilizes the system using $H_\infty$ control design techniques and its outer-loop is used to drive the UAV towards the desired position (Fig. 1). The inner-loop is fast enough to track the given references [28], so that the outer loop dynamics can be approximately described as follows:

$$\dot{x} = u \quad x \in \mathbb{R}^3, \quad u \in U \subseteq \mathbb{R}^3, \tag{1}$$

where $x$ is the position of the UAV; $u$ is the UAV velocity reference generated by the formation algorithm, and $U$ is the velocity constraint set, which is a convex set.

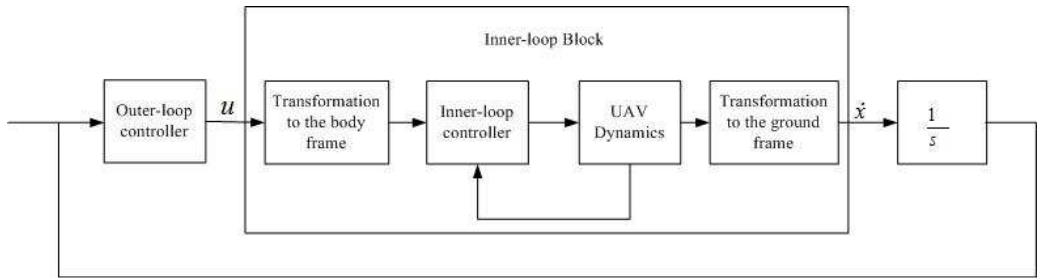

Figure 1: Control Structure of the UAV.

Now, consider the follower velocity in the following form:

$$V_{follower} = V_{leader} + V_{rel}. \tag{2}$$

Having the velocity and position information of the leader, the follower UAV should reach and keep the formation by tuning the relative velocity, $V_{rel}$. Alternatively, one can consider a relatively fixed frame, in which the follower moves with the velocity of $V_{rel}$ and the leader has a relatively fixed position. The problem is to design a supervisor in the outer-loop to bring the follower UAV to the desired distance with respect to the leader position to form the expected formation. Apparently, the desired position moves when the leader changes its position. Here, the control horizon is the sphere, $S_{R_m}$, with the radius of $R_m$ that is centered at the desired position. The formation problem can be stated as follows:

**Problem 1.** *Given the dynamics of the path planner as (1) and the velocity of the follower in the form of (2), design the formation controller to generate the relative velocity of the follower, $V_{rel}$, such that starting from any initial state inside the control horizon, it eventually reaches the desired position, while avoiding the inter-collision between the leader and the follower. Moreover, after reaching the formation, the follower UAV should remain at the desired position for ever.*

To address this problem, we will propose a new method of abstraction based on spherical partitioning and then, as the follower's outer-loop dynamics (1) is a multi-affine function, we will deploy the properties of these classes of functions over the partitioned space. The resulting partitioned system can be bisimilarly captured by a finite DES model for which there are well-established formal analysis and synthesis tools within DES supervisory control framework [30], as described in the next section.

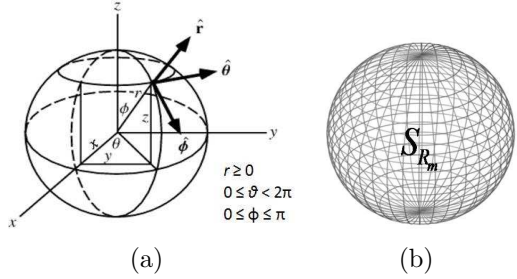

Figure 2: (a) Spherical coordinate system and (b) A partitioned sphere $S_{R_m}$.

## 3. Spherical partitioning

Consider a plant with continuous dynamics of $\dot{x} = g(x) = h(x, u(x))$, defined over the sphere $S_{R_m}$, with the radius of $R_m$, where $u(x)$ is the control value computed based on feedbacked position of the system. The sphere $S_{R_m}$ can be partitioned in the spherical coordinate system with $r \geq 0$, $0 \leq \theta < 2\pi$, and $0 \leq \phi \leq \pi$ (Fig. 2(a)). The curves $\{r = r_i \,|\, 0 \leq r_i \leq R_m, \,for\, i < j : r_i < r_j,\, i,j = 1,...,n_r,\, r_1 = 0,\, r_{n_r} = R_m\}$, $\{\theta = \theta_i \,|\, 0 \leq \theta_i < 2\pi,\, for\, i < j : \theta_i < \theta_j,\, i,j = 1,...,n_\theta,\, \theta_1 = 0,\, \theta_{n_\theta} = 2\pi\}$, and $\{\phi = \phi_i \,|\, 0 \leq \phi_i \leq \pi,\, for\, i < j : \phi_i < \phi_j,\, i,j = 1,...,n_\phi,\, \phi_1 = 0,\, \phi_{n_\phi} = \pi\}$, with $n_r, n_\theta, n_\phi \geq 2$, partition the control horizon $S_{R_m}$. Equivalently partitioning (Fig. 2(b)), we will use $\{r_i = \frac{R_m}{n_r - 1}(i - 1),\, i = 1,...,n_r\}$, $\{\theta_j = \frac{2\pi}{n_\theta - 1}(j - 1),\, j = 1,...,n_\theta\}$, and $\{\phi_k = \frac{\pi}{n_\phi - 1}(k - 1),\, k = 1,...,n_\phi\}$ as partitioning curves. Using this notation, we will have $(n_r - 1)(n_\theta - 1)(n_\phi - 1)$ regions. A region $\bar{R}_{i,j,k} = \{x = (r, \theta, \phi)\,|\, r_i \leq r \leq r_{i+1},\, \theta_j \leq \theta \leq \theta_{j+1},\, \phi_k \leq \phi \leq \phi_{k+1}\}$ is a subset of $S_{R_m}$ surrounded by the above curves. We use the term $R_{i,j,k}$ to denote the interior of the region $\bar{R}_{i,j,k}$.

The intersection between the region $\bar{R}_{i,j,k}$ and the partitioning curves is called a *face* and could be 0-dimensional, 1-dimensional, or 2-dimensional which are named as *vertex*, *edge*, and *facet*, respectively (Fig.3). The set $V(*)$ stands for the vertices that belong to $*$ ($*$ can be a facet, a region $R_{i,j,k}$, or the sphere $S_{R_m}$), and $F(v)$ is the set of facets that the vertex $v$ belongs to them. Each region $R_{i,j,k}$ has eight vertices. The vertices of $R_{i,j,k}$, $V(R_{i,j,k})$, are labeled as follows (Fig.4):

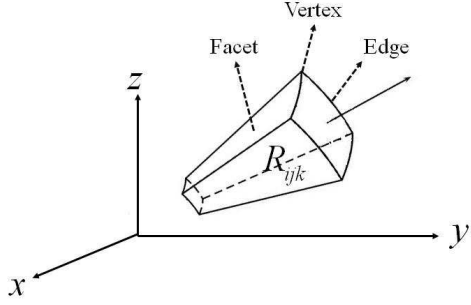

Figure 3: Vertex, Edge, and Facet of the element $R_{i,j,k}$.

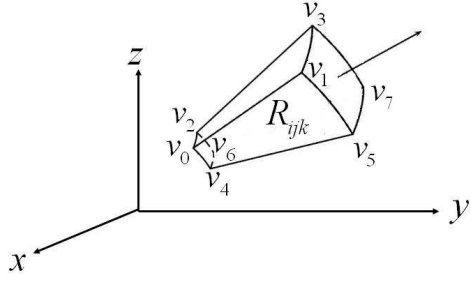

Figure 4: Vertices of the element $R_{i,j,k}$.

$$\begin{cases} v_0 = v_{000} & r = r_i,\ \theta = \theta_j,\ \phi = \phi_k \\ v_1 = v_{001} & r = r_{i+1},\ \theta = \theta_j,\ \phi = \phi_k \\ v_2 = v_{010} & r = r_i,\ \theta = \theta_{j+1},\ \phi = \phi_k \\ v_3 = v_{011} & r = r_{i+1},\ \theta = \theta_{j+1},\ \phi = \phi_k \\ v_4 = v_{100} & r = r_i,\ \theta = \theta_j,\ \phi = \phi_{k+1} \\ v_5 = v_{101} & r = r_{i+1},\ \theta = \theta_j,\ \phi = \phi_{k+1} \\ v_6 = v_{110} & r = r_i,\ \theta = \theta_{j+1},\ \phi = \phi_{k+1} \\ v_7 = v_{111} & r = r_{i+1},\ \theta = \theta_{j+1},\ \phi = \phi_{k+1} \end{cases} \quad (3)$$

We have presented the vertex $v_m$ in the form of $v_{m_\phi,m_\theta,m_r}$, where the binary indices $m_\phi$, $m_\theta$, and $m_r$ show which partitioning curves that have generated the vertex $v_m$. For example, if $m_r = 1$, it shows that the vertex $v_m$ of the region $R_{i,j,k}$ touches the curve $r_{i+1}$, and if $m_r = 0$, it touches the curve $r_i$.

The element $R_{i,j,k}$ has six facets $\{F_r^+, F_r^-, F_\theta^+, F_\theta^-, F_\phi^+, F_\phi^-\}$ and correspond-ingly, six outer normal vectors $\{n_r^+, n_r^-, n_\theta^+, n_\theta^-, n_\phi^+, n_\phi^-\}$ (Fig.5). The exception is when the region $R_{i,j,k}$ touches the origin or the $z$ axis. In this case, some of the vertices are coincident.

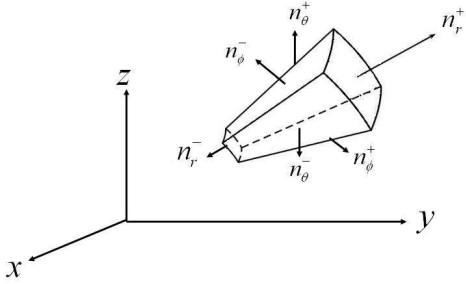

Figure 5: The outer normals of an element $R_{i,j,k}$.

In sphere $S_{R_m}$, let's define $\bar{S}$ as the sphere surface and $E$ as the set of all of the edges and the vertices. Also, consider the detection element $d([i,j,k],[i',j',k']) = \bar{R}_{i,j,k} \cap \bar{R}_{i',j',k'} - E$, which is defined for two regions $R_{i,j,k}$ and $R_{i',j',k'}$ that are adjacent in a common facet (the order is not important). Indeed, the detection elements are the facets in which the edges and the vertices are excluded. With this procedure, the sphere $S_{R_m}$ has been partitioned into $E \cup R_{i,j,k} \cup d([i,j,k],[i',j',k']) \cup S$, where $1 \leq i, i' \leq n_r - 1$, $1 \leq j, j' \leq n_\theta - 1$, $1 \leq k, k' \leq n_\phi - 1$ and $S = \bar{S} - E$. Correspondingly, consider $\tilde{E}$, $\tilde{R}_{i,j,k}$, $\tilde{d}([i,j,k],[i',j',k'])$, $\tilde{S}$ as the labels for these partitioning elements, where $\Im(\tilde{r}) = r$ relates the label $\tilde{r}$ to the set $r$. This partitioned space can be captured by the equivalence relation $Q = \{(x_1, x_2) | \exists \tilde{r} \in \{\tilde{E}, \tilde{R}_{i,j,k}, \tilde{d}([i,j,k],[i',j',k']), \tilde{S}\} \text{ s.t. } x_1, x_2 \in \Im(\tilde{r}) \text{ and } 1 \leq i, i' \leq n_r - 1, 1 \leq j, j' \leq n_\theta - 1, 1 \leq k, k' \leq n_\phi - 1\}$. Correspondingly, the projection map $\pi_Q(x)$ shows the partitioning element that $x$ belongs to it: $\pi_Q(x) = \tilde{r} \in \{\tilde{E}, \tilde{R}_{i,j,k}, \tilde{d}([i,j,k],[i',j',k']), \tilde{S}\}$ s.t. $x \in r$ and $\Im(\tilde{r}) = r$, where $1 \leq i, i' \leq n_r - 1$, $1 \leq j, j' \leq n_\theta - 1$, $1 \leq k, k' \leq n_\phi - 1$.

In the next section, we will utilize the properties of multi-affine functions over the above partitioned space.

## 4. Multi-affine vector fields over spherical partitioned space

Multi-affine functions are a class of functions defined as follows:

**Definition 1.** *Affine function [20]*

A function $g = (g_1, g_2, ..., g_m) : \mathbb{R}^n \to \mathbb{R}^m$ is said to be multi - affine, if all $g_i(x_1, x_2, ..., x_n) : \mathbb{R}^n \to \mathbb{R}$, $i = 1, ..., m$, can be affinely described with respect to each input parameter $x_k$; meaning that if all input parameters are fixed and only $x_k$ changes, then for every set of $a_1, a_2, ..., a_k$ that $\sum_{j=1}^{l} a_j = 1$:

$$g_i(x_1, ..., \sum_{j=1}^{l}(a_j x_{k_j}), ..., x_n) = \sum_{j=1}^{l} a_j g_i(x_1, ..., x_{k_j}, ..., x_n).$$

*4.1. Properties of Multi-affine functions over the partitioned space*

The following proposition shows that a multi-affine function defined over the element $\bar{R}_{i,j,k}$, can be uniquely expressed in terms of its values at the vertices of $R_{i,j,k}$.

**Theorem 1.** *For a multi - affine function $g(x) : S_{R_m} \to \mathbb{R}^3$, the following property holds:*

$$\forall x = (r, \theta, \phi) \in \bar{R}_{i,j,k} : g(x) = \sum_m \lambda_m g(v_m), \quad m = 0, 2, ..., 7, \qquad (4)$$

*where $v_m$, $m = 0, ..., 7$, are the vertices of the element $R_{i,j,k}$ and $\lambda_m$ can be obtained uniquely as follows:*

$$\lambda_m = \lambda_r^{m_r}(1 - \lambda_r)^{1-m_r} \lambda_\theta^{m_\theta}(1 - \lambda_\theta)^{1-m_\theta} \lambda_\phi^{m_\phi}(1 - \lambda_\phi)^{1-m_\phi}, \qquad (5)$$

*where $m_r$, $m_\theta$, $m_\phi$ are corresponding binary digits of the index $m$ as declared in (3) and*

$$\lambda_r = \frac{r - r_i}{r_{i+1} - r_i} \qquad \lambda_\theta = \frac{\theta - \theta_j}{\theta_{j+1} - \theta_j} \qquad \lambda_\phi = \frac{\phi - \phi_k}{\phi_{k+1} - \phi_k}$$

*proof*: See the Appendix for the proof. ∎

**Remark 1.** *It can be verified that the resulting coefficients $\lambda_m$ has the property that $\lambda_m \geq 0$ and $\sum_m \lambda_m = 1$, $m = 0, 1, ..., 7$.*

Theorem 1 also holds true for the points on the facets as described in the next proposition:

**Proposition 1.** *For a multi - affine function defined over $\bar{R}_{i,j,k}$, for all of the facets $F_q^s$ of $R_{i,j,k}$, $q \in \{r, \theta, \phi\}$ and $s \in \{+, -\}$, the following property*

holds:

$$\forall x = (r, \theta, \phi) \in F_q^s : g(x) = \sum_{v_m} \lambda_m g(v_m), \ v_m \in V(F_q^s). \quad (6)$$

*proof*: It is a special case of Theorem 1 and, to prove this proposition, it just needs to follow only steps 1 and 2 in the proof of Theorem 1. ∎

In the following proposition, we will show that in Theorem 1, the coefficients are unique and there is one and only one multi-affine function over the sphere $S_{R_m}$ that takes fixed values like $g(v_m)$ at the vertices of $R_{i,j,k}$.

**Proposition 2.** *Consider a map $g : v(S_{R_m}) \to \mathbb{R}^3$. Then, there exists one and only one multi-affine function $f : S_{R_m} \to \mathbb{R}^3$ defined over the region $R_{i,j,k}$ with the vertices $V_m$, $m = 0, ..., 7$, that satisfies $f(v_m) = g(v_m)$, for all of the vertices.*

*Proof*: See the Appendix for proof. ∎

*4.2. Utilizing the Multi-affine functions over the partitioned space*

For a multi-affine vector field defined over the region $R_{i,j,k}$, two important control features can be defined: *the invariant region* in which the system trajectories do not leave the region (Fig.6(a)) and *the exit facet* through which the trajectories of the system leave the region (Fig 6(b)). The formal definition is given as follows:

**Definition 2. *Invariant region***
In the sphere $S_{R_m}$ and the multi-affine vector field $\dot{x} = g(x)$, $g : S_{R_m} \to \mathbb{R}^3$, the region $R_{i,j,k}$ is said to be an invariant region, if

$$\forall x(0) \in R_{i,j,k} \Rightarrow x(t) \in R_{i,j,k} \ for \ t \geq 0$$

**Definition 3. *Exit facet***
In the sphere $S_{R_m}$ and the multi-affine vector field $\dot{x} = g(x)$, $g : S_{R_m} \to \mathbb{R}^3$, the facet $F_q^s$, $q = \{r, \theta, \phi\}$, and $s = \{+, -\}$ is an exit facet, if $\forall x(0) \in R_{i,j,k} \Rightarrow \exists \tau \ (finite) \geq 0 \ and \ \exists \varepsilon \geq 0 \ satisfying$:

1. $x(t) \in R_{i,j,k} \quad for \ t \in [0, \tau)$
2. $x(t) \in F_q^s \quad \quad for \ t = \tau$
3. $x(t) \notin \bar{R}_{i,j,k} \quad for \ t \in (\tau, \tau + \varepsilon)$

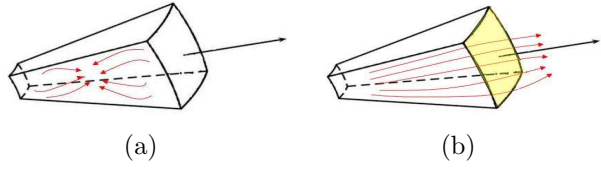

Figure 6: (a) Invariant region and (b) Exit facet.

In the following theorems, we will find what conditions are required to construct a feedback controller to have an invariant region $R_{i,j,k}$, or an exit facet $F_q^s$.

**Theorem 2.** *Sufficient condition for $R_{i,j,k}$ to be an invariant region:* *For a multi-affine vector field $\dot{x} = g(x)$, $g : S_{R_m} \to \mathbb{R}^3$, $R_{i,j,k}$ is an invariant region if for each facet $F_q^s$ and its corresponding outer normal $n_q^s$, $q \in \{r, \theta, \phi\}$ and $s \in \{+, -\}$:*

$$n_q^s(y)^T . g(v_m) < 0 \; \forall v_m \in V(F_q^s), \; \forall y \in F_q^s \qquad (7)$$

*proof*: We should show that $n_q^s(y)^T . g(y) < 0$ holds true for any $y \in F_q^s$ and for all $F_q^s$. First, according to Proposition 1, we know that:

$$\forall y \in F_q^s : g(y) = \sum_{v_m} \lambda_m g(v_m), \; v_m \in V(F_q^s) \qquad (8)$$

Using this value of $g(y)$, we can write:

$$n_q^s(y)^T . g(y) = n_q^s(y)^T . \sum_{v_m} \lambda_m g(v_m) = \sum_{v_m} \lambda_m \, n_q^s(y)^T . g(v_m) \qquad (9)$$

Now, according to the problem assumption described in (7), we know that $n_q^s(y)^T . g(v_m) < 0$ for all $v_m \in V(F_q^s)$ and $y \in F_q^s$. On the other hand, according to Remark 1, we have $\lambda_m \geq 0$ and $\sum_m \lambda_m = 1$. Hence, from (9), it can be concluded that $n_q^s(y)^T . g(y) < 0$:

This means that the trajectories of the system cannot leave $R_{i,j,k}$ through the facet $F_q^s$. Since this is through for all of the facets, the trajectories of the system will not leave the region $R_{i,j,k}$ and will remain inside it forever. ■

**Theorem 3.** *Sufficient condition for an Exit facet:* *For a multi-affine vector field $\dot{x} = g(x)$, $g : S_{R_m} \to \mathbb{R}^3$, the facet $F_q^s$ with the outer normal $n_q^s$, $q \in \{r, \theta, \phi\}$ and $s \in \{+, -\}$, is an exit facet if:*

1. $n_{q'}^{s'}(y)^T.g(v_m) < 0 \quad \forall v_m \in V(F_{q'}^{s'}), \forall y \in F_{q'}^{s'}, q' \neq q, \text{ or } s' \neq s$
2. $n_q^s(y)^T.g(v_m) > 0 \quad \forall v_m \in V(R_{i,j,k}), \text{ for all } y \in F_q^s$

*proof*: The first requirement guarantees that the trajectories of the system do not leave $R_{i,j,k}$ through the non-exit facets $F_{q'}^{s'} \neq F_q^s$. This has already been proven in Theorem 2. The second requirement is to drive the trajectory of the system out through the facet $F_q^s$. Based on the assumption, for all $y \in F_q^s$ and for all $v_m \in V(R_{i,j,k})$, we have: $n_q^s(y)^T.g(v_m) > 0$. According to Theorem 1, for the multi-affine function $g$, there exist $\lambda_m$ such that $\forall x \in \bar{R}_{i,j,k} : g(x) = \sum_m \lambda_m g(v_m), \ m = 0, 1, ..., 7$. Since $\lambda_m \geq 0$ and $\sum_m \lambda_m = 1$, then $n_q^s(y)^T.\lambda_m g(v_m) > 0 \ for \ all \ v_m$. This will lead to have $n_q^s(y)^T.g(x) > 0$ for all $x \in \bar{R}_{i,j,k}$, which means that the trajectories of the system have a strictly positive velocity in the direction of $n_q^s(y)$ steering them to exit from $R_{i,j,k}$ through the facet $F_q^s$. ∎

Here, we will present some of the properties of the properties of the exit facet controller that will be used through our further derivations.

**Lemma 1.** *For a multi-affine vector field $\dot{x} = g(x)$, $g : S_{R_m} \to \mathbb{R}^3$, in a region $R_{i,j,k}$ with the exit facet $F_q^s$, constructed by Theorem 3, the following properties are concluded:*

1. *The trajectories that leave the region do not return back any more.*
2. *The points on the exit facet are not reachable from other points on the facet.*
3. *The trajectory that has reached the exit facet, leaves it immediately.*

*proof*: As observed in the proof of Theorem 3, respecting the second condition of this theorem leads to have $n_q^s(y)^T.g(x) > 0$ for any $x \in \bar{R}_{i,j,k}$ and any $y \in F_q^s$. In particular this is true for the points on the facet $F_q^s$. Hence, $n_q^s(y)^T.g(y) > 0, \forall y \in F_q^s$. This strictly positive inequality guarantees that the trajectory that reaches the exit facet, leaves it upon reaching the facet so that it can neither move along the facet nor return back. ∎

**Lemma 2.** *For a multi-affine vector field $\dot{x} = g(x)$, $g : S_{R_m} \to \mathbb{R}^3$, in a region $R_{i,j,k}$ with the exit facet $F_q^s$ constructed by Theorem 3, the trajectories that leave the region only can pass through the detection elements.*

*proof*: As we saw in the proof of Theorem 2, for all points on the non-exit facets $F_{q'}^{s'}$, we have $n_{q'}^{s'}(y)^T.g(y) < 0$. This strictly negative inequality shows that the trajectories of the system cannot pass through the non-exit facets including the edges and the vertices that belong to them. In particular, the trajectories cannot cross the edges and the vertices that are common between the non-exit facets and the exit facet. On the other hand, Theorem 3 shows that the trajectories of the system cannot remain inside the region. Hence, the only way is that the trajectories pass through the the internal area of the exit facet, which we have called it as the detection element. ∎

**Proposition 3.** *For a multi-affine vector field $\dot{x} = g(x)$, $g : S_{R_m} \to \mathbb{R}^3$, in a region $R_{i,j,k}$ with the exit facet $F_q^s$ constructed by Theorem 3, all $y \in F_q^s \setminus E$ are reachable from a point inside the region $R_{i,j,k}$.*

*proof*: Since any $y \in F_q^s$ is not reachable form the adjacent region (Part 1, Lemma 1) or from another point on $F_q^s$ (Part 2, Lemma 1), then, considering $n_q^s(y)^T.g(y) > 0$, by continuity of $g$, there is a point inside the region $R_{i,j,k}$ on the neighborhood of $y$ from which $y$ is reachable. ∎

## 5. Control over the partitioned space

### 5.1. Eligible sets for Invariant region

Consider the multi-affine vector field $\dot{x} = g(x) = h(x, u(x))$ over the partitioned space. Following from Theorem 2, to construct the region $R_{i,j,k}$ as an invariant region, it is required to satisfy inequality (7) for all of the facets of the region. Starting from $F_r^+$, we then need to have :

$$n_r^{+^T}.g(v_m) < 0 \quad for \ \ v_m \in V(F_r^+) \tag{10}$$

where:
$$n_r^+ = (1, \theta, \phi) \quad \theta_j \leq \theta \leq \theta_{j+1} \ and \ \phi_k \leq \phi \leq \phi_{k+1} \tag{11}$$

Solving these inequalities is not an easy job. In [26] a geometric way is introduced to solve these inequalities in a polar coordinate system. To satisfy (10), extending this method to the spherical coordinates, the value of $g(v_m)$ should be chosen from a set $g_{F_r^+}(R_{i,j,k})$, for all $v_m \in V(F_r^+)$, where $g_{F_r^+}(R_{i,j,k}) = \{(r, \theta, \phi) | \theta_{j+1} + \frac{\pi}{2} \leq \theta \leq \theta_j + \frac{3\pi}{2} \ and \ \phi_{k+1} + \frac{\pi}{2} \leq \phi \leq \phi_k + \frac{3\pi}{2}\}$. As $\theta$ and $\phi$ must be within certain ranges: $0 \leq \theta < 2\pi$, $0 \leq \phi \leq \pi$, we define the following functions:

$$\hbar(\alpha) = \begin{cases} 1 & \text{for} \quad 2k\pi \leq \alpha \leq (2k+1)\pi \\ -1 & \text{for} \ (2k+1)\pi \leq \alpha < (2k+2)\pi \end{cases}$$

and

$$range(r, \theta, \phi) = (r, \theta', \phi'), \text{ where}$$

$$\theta' = \begin{cases} \theta - 2k\pi & \text{for} \ \hbar(\theta)\hbar(\phi) > 0, \ 2k\pi \leq \theta \leq (2k+2)\pi \\ \theta - (2k-1)\pi & \text{for} \ \hbar(\theta)\hbar(\phi) < 0, \ 2k\pi \leq \theta \leq (2k+1)\pi \\ \theta - (2k+1)\pi & \text{for} \ \hbar(\theta)\hbar(\phi) < 0, \ (2k+1)\pi < \theta < (2k+2)\pi \end{cases}$$

and

$$\phi' = \begin{cases} \phi - 2k\pi & \text{for} \ 2k\pi \leq \phi \leq (2k+1)\pi \\ (2k+2)\pi - \phi & \text{for} \ (2k+1)\pi \leq \phi \leq (2k+2)\pi \end{cases}$$

Using these functions, the eligible facet sets for $g_{F_r^+}$ and other facets can be obtained as follows:

$$\begin{cases} g_{F_r^+}(R_{i,j,k}) = \{range(r, \theta, \phi) | \theta_{j+1} + \frac{\pi}{2} \leq \theta \leq \theta_j + \frac{3\pi}{2} \text{ and} \\ \qquad \phi_{k+1} + \frac{\pi}{2} \leq \phi \leq \phi_k + \frac{3\pi}{2}\} \\ g_{F_r^-}(R_{i,j,k}) = \{range(r, \theta, \phi) | \theta_{j+1} - \frac{\pi}{2} \leq \theta \leq \theta_j + \frac{\pi}{2} \text{ and} \\ \qquad \phi_{k+1} - \frac{\pi}{2} \leq \phi \leq \phi_k + \frac{\pi}{2}\} \\ g_{F_\theta^+}(R_{i,j,k}) = \{range(r, \theta, \phi) | \theta_{j+1} - \pi \leq \theta \leq \theta_{j+1}\} \\ g_{F_\theta^-}(R_{i,j,k}) = \{range(r, \theta, \phi) | \theta_j \leq \theta \leq \theta_j + \pi\} \\ g_{F_\phi^+}(R_{i,j,k}) = \{range(r, \theta, \phi) | \phi_{k+1} - \pi \leq \phi \leq \phi_{k+1}\} \\ g_{F_\phi^-}(R_{i,j,k}) = \{range(r, \theta, \phi) | \phi_k \leq \phi \leq \phi_k + \pi\} \end{cases} \qquad (12)$$

Now, since $v_0 \in V(F_r^-) \cap V(F_\theta^-) \cap V(F_\phi^-)$, to satisfy (7), the value of $g(v_0)$ should be chosen from the set $Inv_0 = g_{F_r^-} \cap g_{F_\theta^-} \cap g_{F_\phi^-}$. The same procedure should be followed for other vertices. The eligible vertex sets for the region $R_{i,j,k}$, are as follows:

$$\begin{cases}
Inv_0(R_{i,j,k}) = \{range(r,\theta,\phi)|\, \theta_j < \theta < \theta_j + \frac{\pi}{2} \text{ and } \phi_k < \phi < \phi_k + \frac{\pi}{2}\} \\
Inv_1(R_{i,j,k}) = \{range(r,\theta,\phi)|\, \theta_{j+1} + \frac{\pi}{2} < \theta < \theta_j + \pi \text{ and} \\
\quad \phi_{k+1} + \frac{\pi}{2} < \phi < \phi_k + \pi\} \\
Inv_2(R_{i,j,k}) = \{range(r,\theta,\phi)|\, \theta_{j+1} - \frac{\pi}{2} < \theta < \theta_{j+1} \text{ and } \phi_k < \phi < \phi_k + \frac{\pi}{2}\} \\
Inv_3(R_{i,j,k}) = \{range(r,\theta,\phi)|\, \theta_{j+1} - \pi < \theta < \theta_j + \frac{3\pi}{2} \text{ and} \\
\quad \phi_{k+1} + \frac{\pi}{2} < \phi < \phi_k + \pi\} \\
Inv_4(R_{i,j,k}) = \{range(r,\theta,\phi)|\, \theta_j < \theta < \theta_j + \frac{\pi}{2} \text{ and } \phi_{k+1} - \frac{\pi}{2} < \phi < \phi_{k+1}\} \\
Inv_5(R_{i,j,k}) = \{range(r,\theta,\phi)|\, \theta_{j+1} + \frac{\pi}{2} < \theta < \theta_j + \pi \text{ and} \\
\quad \phi_{k+1} + \pi < \phi < \phi_k + \frac{3\pi}{2}\} \\
Inv_6(R_{i,j,k}) = \{range(r,\theta,\phi)|\, \theta_{j+1} - \frac{\pi}{2} < \theta < \theta_{j+1} \text{ and} \\
\quad \phi_{k+1} - \frac{\pi}{2} < \phi < \phi_{k+1}\} \\
Inv_7(R_{i,j,k}) = \{range(r,\theta,\phi)|\, \theta_{j+1} + \pi < \theta < \theta_j + \frac{3\pi}{2} \text{ and} \\
\quad \phi_{k+1} + \pi < \phi < \phi_k + \frac{3\pi}{2}\}
\end{cases} \quad (13)$$

*5.2. Eligible sets for Exit facets*

Assume that in the region $R_{i,j,k}$, we are going to construct $F_r^-$ as an exit facet. Following from Theorem 3, for all of the vertices, $v_m$, $m = 0, ..., 7$, two conditions should be checked. Let's start with $v_0$. As $v_0$ belongs to the facets $F_r^-$, $F_\theta^-$, $F_\phi^-$, and $F_r^-$ is the exit facet, to satisfy the first condition of Theorem 3, it is sufficient to find $g(v_0)$ such that $n_\theta^-(x)^T.g(v_0) < 0$ for all $x \in F_\theta^-$ and $n_\phi^-(x)^T.g(v_0) < 0$ for all $x \in F_\phi^-$. To satisfy these inequalities, it needs to have $g(v_0) \in g_{F_\phi^-} \cap g_{F_\theta^-}$.

To satisfy the second condition of Theorem 3, it needs to find $g(v_m)$, $m = 0, ..., 7$, such that they satisfy $n_r^-(y)^T.g(v_m) > 0$ for all $y \in F_r^-$. In this case, the value of $g(v_0)$ can be chosen from the set $h_{F_r^-} = \{g|n_r^-(x)^T.g > 0 \,\forall\, x \in F_r^-\}$.

To satisfy both conditions, $g(v_0)$ should be chosen from the eligible exit facet set $Ex_0(F_r^-(R_{i,j,k})) = h_{F_r^-} \cap g_{F_\phi^-} \cap g_{F_\theta^-}$. This procedure can be similarly continued to find the eligible sets for other vertices, $Ex_m(F_r^-(R_{i,j,k}))$, $m = 0, ..., 7$. Following this procedure, to have $F_r^-$ as an exit facet of the region $R_{i,j,k}$, the eligible exit facet sets are:

$$\begin{cases} Ex_0(F_r^-(R_{i,j,k})) = Ex_1(F_r^-(R_{i,j,k})) = \\ \quad \{range(r,\theta,\phi)|\, \theta_{j+1} + \frac{\pi}{2} < \theta < \theta_j + \pi \text{ and } \phi_{k+1} + \frac{\pi}{2} < \phi < \phi_k + \pi\} \\ Ex_2(F_r^-(R_{i,j,k})) = Ex_3(F_r^-(R_{i,j,k})) = \\ \quad \{range(r,\theta,\phi)|\, \theta_{j+1} + \frac{\pi}{2} < \theta < \theta_j + \pi \text{ and } \phi_{k+1} + \pi < \phi < \phi_k + \frac{3\pi}{2}\} \\ Ex_4(F_r^-(R_{i,j,k})) = Ex_5(F_r^-(R_{i,j,k})) = \\ \quad \{range(r,\theta,\phi)|\, \theta_{j+1} + \pi < \theta < \theta_j + \frac{3\pi}{2} \text{ and } \phi_{k+1} + \frac{\pi}{2} < \phi < \phi_k + \pi\} \\ Ex_6(F_r^-(R_{i,j,k})) = Ex_7(F_r^-(R_{i,j,k})) = \\ \quad \{range(r,\theta,\phi)|\, \theta_j + \pi < \theta < \theta_j + \frac{3\pi}{2} \text{ and } \phi_{k+1} + \pi < \phi < \phi_k + \frac{3\pi}{2}\} \end{cases} \quad (14)$$

The procedure that makes other facets as exit facet is similar.

*5.3. Construction of the controllers*

By proper selection of the control value $u(x)$, it is possible to tune the multi-affine vector field, $\dot{x} = g(x) = h(x, u(x))$, at the vertices, so that the region $R_{i,j,k}$ becomes an invariant region or one of its facets becomes an exit facet. Indeed, to make the region $R_{i,j,k}$ as an invariant region, the value of the control signal at the vertices, $u(v_m)$, should be chosen such that $g(v_m) = h(v_m, u(v_m))$ falls in the set $U_m(Inv(R_{i,j,k})) = Inv_m(R_{i,j,k}) \cap U$, for $m = 0, ..., 7$, where $Inv_m(R_{i,j,k})$ is the eligible set for the vertex $v_m$ as introduced in (13) and $U$ is the velocity bound, which comes from the practical limitations. If $U_m(Inv(R_{i,j,k})) \neq \emptyset$, for all $m = 0, ..., 7$, then making the region $R_{i,j,k}$ as an invariant region is feasible. Based on Theorem 1, having the value of the control function $u$ at the vertices $v_m$, it is possible to construct the multi-affine controller $u(x)$ for all $x \in \bar{R}_{i,j,k}$. Later, we will use the notation $C_0(R_{i,j,k})$ to label this controller. In the case that for all of the regions $R_{i,j,k}$, the values of the vector field at the vertices, $u(v_m)$, are the same, then we can simply use the label $C_0$ to denote the controller $C_0(R_{i,j,k})$.

To make the facet $F_q^s$ as an exit facet, similar to the invariant controller, it is sufficient to choose the values of $u(v_m)$ such that $g(v_m) = h(v_m, u(v_m))$ falls in the set $U_m(Ex(F_q^s(R_{i,j,k}))) = Ex_m(F_q^s(R_{i,j,k})) \cap U$. Therefore, for a region $R_{i,j,k}$, if all of $U_m(Ex(F_q^s(R_{i,j,k}))) \neq \emptyset$, then corresponding to each of the exit facets, $F_r^+$, $F_r^-$, $F_\theta^+$, $F_\theta^-$, $F_\phi^+$, $F_\phi^-$, there are controllers that can make them as exit facets. We label these controllers as $C_r^+$, $C_r^-$, $C_\theta^+$, $C_\theta^-$, $C_\phi^+$, $C_\phi^-$, respectively.

**Remark 2.** *For a convex velocity constraint set, $U$, since $g(x)$ is a multi-affine function, respecting the velocity bounds at the vertices, leads the system to respect these conditions for all $x \in \bar{R}_{i,j,k}$.*

## 6. Spherical abstraction of the state space

Now, consider the original system which is defined over the partitioned space. The equivalence relation $Q$, defined in Section 3, describes this partitioned space. This system can be captured by the transition system $T_Q = (X_Q, X_{Q_0}, U_Q, \rightarrow_Q, Y_Q, H_Q)$, where

- $X_Q = E \cup R_{i,j,k} \cup d([i,j,k],[i',j',k']) \cup S$ is the set of system states, where $1 \le i, i' \le n_r - 1$, $1 \le j, j' \le n_\theta - 1$, $1 \le k, k' \le n_\phi - 1$.

- $X_{Q_0}$ is the set of initial states. Assuming that the system initially starts from inside one of the regions $R_{i,j,k}$, $X_{Q_0} = \bigcup R_{i,j,k}$, where $1 \le i, i' \le n_r - 1$, $1 \le j, j' \le n_\theta - 1$, $1 \le k, k' \le n_\phi - 1$.

- $U_Q = U_a \cup U_d$, where

    - $U_a = \{C_r^+, C_r^-, C_\theta^+, C_\theta^-, C_\phi^+, C_\phi^-, C_0\}$ is the set of labels corresponding to the controllers that can make the region $R_{i,j,k}$ as an invariant region or can make one of its facets as an exit facet. For these control labels, the sets of control actions that can be activated in this region are: $r(C_q^s) = \{u(x)|u(x) = \sum_m \lambda_m u(v_m), m = 0, 1, ..., 7, v_m \in V(R_{i,j,k}), u(v_m) \in U_m(Ex(F_q^s))\}$, and $r(C_0) = \{u(x)|u(x) = \sum \lambda_m u(v_m), v_m \in V(R_{i,j,k}), u(v_m) \in U_m(Inv(R_{i,j,k}))\}$. $\lambda_m$ can be obtained by (5).

    - $U_d = U_c \cup U_e$ is the set of the detection events, where $U_c = \{\hat{d}([i,j,k],[i',j',k'])|\, 1 \le i, i' \le n_r - 1,\ 1 \le j, j' \le n_\theta - 1,\ 1 \le k, k' \le n_\phi - 1\}$. $\hat{d}([i,j,k],[i',j',k'])$ is an event that shows the detection element $d([i,j,k],[i',j',k'])$ has been crossed and $U_e$ is the set of external events such as entering into an alarm zone of collision.

- $(x, x', v) \in \rightarrow_Q$, denoted by $x \xrightarrow{v}_Q x'$, if and only if one of the following conditions holds true:

    1. Actuation:

- Exit facet: $v \in \{C_q^s | q \in \{r, \theta, \phi\}, s \in \{+, -\}\}$; $\pi_Q(x) \neq \pi_Q(x')$; $\pi_Q(x) = \tilde{R}_{i,j,k}$; $\pi_Q(x') = \tilde{d}([i, j, k], [i', j', k'])$; $\exists \tau (finite)$ and $\varepsilon > 0$ s.t. $\psi(t) : [0, \tau+\varepsilon] \to \mathbb{R}^3$ is the solution of $\dot{x} = h(x, r(v))$; $\psi(0) = x$; $\psi(\tau) = x'$; $\pi_Q(\psi(t)) = \pi_Q(x)$ for $t \in [0, \tau)$, and $\pi_Q(\psi(t)) \neq \pi_Q(x)$ for $t \in [\tau, \tau + \varepsilon]$. Here, $r(v)$ is the continuous controller corresponding to the the control label $v$, which can be constructed as discussed in Section 5.3.
- Invariant region: $v = C_0$; $\pi_Q(x) = \pi_Q(x') = \tilde{R}_{i,j,k}$; $\psi(t) : \mathbb{R}^+ \to \mathbb{R}^3$ is the solution of $\dot{x} = h(x, r(v))$; $\psi(0) = x$; $\psi(\tau) = x'$, and $\pi_Q(\psi(t)) = \pi_Q(x)$ $for\ all\ t \geq 0$.

2. detection:
   - Crossing a detection element: $v \in U_d$; $\pi_Q(x) \neq \pi_Q(x')$; $\pi_Q(x) = \tilde{d}([i, j, k], [i', j', k'])$; $\pi_Q(x') = \tilde{R}_{i',j',k'}$; $\exists 0 < \varepsilon < \tau$ and $\exists w \in \{C_q^s | q \in \{r, \theta, \phi\}, s \in \{+, -\}\}$ s.t. $\psi(t) : [0, \tau] \to \mathbb{R}^3$ is the solution of $\dot{x} = h(x, r(w))$; $\psi(\varepsilon) = x$; $\psi(\tau) = x'$; $\pi_Q(\psi(t)) = \tilde{R}_{i,j,k}$ $for\ t \in (0, \varepsilon)$, and $\pi_Q(\psi(t)) = \tilde{R}_{i',j',k'}$ $for\ t \in (\varepsilon, \tau]$.
   - External events: $v \in U_e$, and $x = x'$. In this case, $x$ is the value of the system state at the time instant that the event $v$ appears. The external event does not affect the system dynamics.

- $Y_Q = X_Q$ is the output space.

- $H_Q : X \to Y_Q$ is the output map. Here, we have chosen $H_Q(x) = \pi_Q(x)$.

Although $T_Q$ contains only important transitions that either cross the boundaries or remain inside the regions, still it has infinite states and the analysis of such a system might be difficult. Abstraction [31], is the technique that can reduce the complexity and can lead to a finite state machine for which the DES supervisory control tools can be used for the system analysis and control synthesis. To do so, each partitioning element, can be considered one state in the abstracted model. Hence, the abstract model is a tuple $T_\xi = (X_\xi, X_{\xi_0}, U_\xi, \to_\xi, Y_\xi, H_\xi)$, where

- $X_\xi = \{\tilde{R}_{i,j,k} | 1 \leq i \leq n_r - 1,\ 1 \leq j \leq n_\theta - 1,\ 1 \leq k \leq n_\phi - 1\} \bigcup \{\tilde{d}([i, j, k], [i', j', k']) | 1 \leq i, i' \leq n_r - 1,\ 1 \leq j, j' \leq n_\theta - 1,\ 1 \leq k, k' \leq n_\phi - 1\}$, where $\tilde{R}_{i,j,k}$ and $\tilde{d}([i, j, k], [i', j', k'])$ are the labels for the regions $R_{i,j,k}$ and $d([i, j, k], [i', j', k'])$, respectively. Note that since

the system starts from a point inside the regions $R_{i,j,k}$ and never crosses the edges or the vertices, the set $E$ does not need to be considered in the abstracted system. Moreover, as the sphere $S_{R_m}$ is the control horizon, its surface, $S$, should not be crossed.

- $X_{\xi_0} = \{\tilde{R}_{i,j,k}|\ 1 \leq i, i' \leq n_r-1,\ 1 \leq j, j' \leq n_\theta-1,\ 1 \leq k, k' \leq n_\phi-1\}$.

- $U_\xi = U_a \cup U_d$ is like what we have in $T_Q$.

- $(r, r', v) \in \to_\xi$, denoted by $r \xrightarrow{v}_\xi r'$, if $\exists v \in U_\xi$, $x \in \Im(r)$, $x' \in \Im(r')$ such that $x \xrightarrow{v}_Q x'$.

- $Y_\xi = X_\xi$.

- $H_\xi(r) = r$ is the output map, which is selected as an identity map.

In general, the abstract model contains all of the behaviors of the partitioned system, however, the converse might not be always true. If the converse is also true, we say that they are bisimilar. A bisimulation relation between two transition systems can be formally defined as follows:

**Definition 4.** *[31] Given $T_i = (Q_i, Q_i^0, U_i, \to_i, Y_i, H_i)$, $(i = 1, 2)$, $R$ is a bisimulation relation between $T_1$ and $T_2$, denoted by $T_1 \approx T_2$, iff:*

1. $\forall q_1 \in Q_1^0$ then $\exists q_2 \in Q_2^0$ that $(q_1, q_2) \in R$ and $\forall q_2 \in Q_2^0$ then $\exists q_1 \in Q_1^0$ that $(q_1, q_2) \in R$.
2. $\forall q_1 \to_1 q_1'$, and $(q_1, q_2) \in R$ then $\exists q_2' \in Q_2$ such that $q_2 \to_2 q_2'$ and $(q_1', q_2') \in R$. Also, $\forall q_2 \to_2 q_2'$, and $(q_1, q_2) \in R$ then $\exists q_1' \in Q_1$ such that $q_1 \to_1 q_1'$ and $(q_1', q_2') \in R$.

**Theorem 4.** *The original partitioned system, $T_Q$, and the abstract model, $T_\xi$, are bisimilar.*

*Proof*: Consider the relation $R = \{(q_Q, q_\xi)|q_Q \in X_Q,\ q_\xi \in X_\xi, and\ q_Q \in \Im(q_\xi)\}$. We will show that this relation is a bisimulation relation between $T_Q$ and $T_\xi$.

To verify the first condition of bisimulation relation, for any $q_Q \in X_{Q_0}$ there exists a region $R_{i,j,k}$ such that $q_Q \in R_{i,j,k}$. For this region, there exists a label, $\tilde{R}_{i,j,k}$ such that $R_{i,j,k} = \Im(\tilde{R}_{i,j,k})$ and $\tilde{R}_{i,j,k} \in X_{\xi_0}$. Hence, $(q_Q, \tilde{R}_{i,j,k}) \in R$. Conversely, it can be similarly shown that for any $q_\xi \in X_{\xi_0}$, there exists a $q_Q \in X_{Q_0}$ such that $(q_\xi, q_Q) \in R$.

For the second condition of bisimulation relation, following from the definition of $T_\xi$, for any $(q_Q, q_\xi) \in R$ and $q_Q \xrightarrow{u}_Q q'_Q$, there exists a transition $q_\xi \xrightarrow{u}_\xi q'_\xi$, where $q'_Q \in \Im(q'_\xi)$, or equivalently $(q'_Q, q'_\xi) \in R$. For the converse case, assume that $q_\xi \xrightarrow{u}_\xi q'_\xi$. According to the definition of $R$, all $x \in \Im(q_\xi)$ are related to $q_\xi$. Hence, to prove the second condition of the bisimulation relation, we should investigate it for all $x \in \Im(q_\xi)$. Based on the control construction procedure, the labels $u$, $q_\xi$, and $q'_\xi$ can be one of the following cases:

1. $u = C_0$ and $q_\xi = q'_\xi$. In this case, since the controller $C_0$ makes the region as an invariant region (Theorem 2), all of the trajectories starting from any $q_Q \in \Im(q_\xi)$ will remain inside the region $\Im(q_\xi)$. Therefore, for any $q_Q \in \Im(q_\xi)$, there exists a $q'_Q \in \Im(q_\xi)$ such that $q_Q \xrightarrow{u}_Q q'_Q$ and $q'_Q = \Im(q'_\xi)$.

2. $u \in C_q^s$, $q_\xi \in \{\tilde{R}_{i,j,k} | 1 \leq i \leq n_r - 1,\ 1 \leq j \leq n_\theta - 1,\ 1 \leq k \leq n_\phi - 1\}$, and $q'_\xi \in \{\tilde{d}([i,j,k], [i',j',k']) | 1 \leq i, i' \leq n_r - 1,\ 1 \leq j, j' \leq n_\theta - 1,\ 1 \leq k, k' \leq n_\phi - 1\}$. In this case, based on Theorem 3 and Lemma 2, starting from any $q_Q \in \Im(q_\xi)$, the controller $C_q^s$ drives the system trajectory towards the detection element $\Im(q'_\xi)$. Therefore, for any $q_Q \in \Im(q_\xi)$, there exists a $q'_Q \in \Im(q'_\xi)$ such that $q_Q \xrightarrow{u}_Q q'_Q$ and $q'_Q \in \Im(q'_\xi)$.

3. $u \in U_c = \{\hat{d}([i,j,k], [i',j',k']) | 1 \leq i, i' \leq n_r - 1,\ 1 \leq j, j' \leq n_\theta - 1,\ 1 \leq k, k' \leq n_\phi - 1\}$ and $q'_\xi \in \{\tilde{R}_{i',j',k'} | 1 \leq i' \leq n_r - 1,\ 1 \leq j' \leq n_\theta - 1,\ 1 \leq k' \leq n_\phi - 1\}$, and $q_\xi \in \{\tilde{d}([i,j,k], [i',j',k']) | 1 \leq i, i' \leq n_r - 1,\ 1 \leq j, j' \leq n_\theta - 1,\ 1 \leq k, k' \leq n_\phi - 1\}$. In this case, based on Proposition 3, for any $q_Q \in \Im(q_\xi)$ there is a controller $v \in C_q^s$ that has leads the trajectory of the system from the region $R_{i,j,k}$ to the point $q_Q$ on the detection element $d([i,j,k], [i',j',k'])$. Since $R_{i',j',k'}$ is the unique adjacent region of the element $R_{i,j,k}$ common in the detection element $d([i,j,k], [i',j',k'])$, based on the definition of the controller for the exit facet and Theorem 3, the controller $v$ leads the trajectory of the system to a point inside the region $R_{i',j',k'}$ so that the detection event $u = \hat{d}([i,j,k], [i',j',k'])$ is generated. Therefore, for any $q_Q \in \Im(q_\xi)$, there exists a $q'_Q \in \Im(q'_\xi)$ such that $q_Q \xrightarrow{u}_Q q'_Q$ and $q'_Q \in \Im(q'_\xi)$.

4. $u \in U_e$ is the external event. In this case, the state of system does not change, meaning that $q_Q = q'_Q$ and $q_\xi \in q'_\xi$. Therefore, trivially for any $q_Q \in \Im(q_\xi)$ and $q_\xi \to_\xi q'_\xi$, we have $q_Q \xrightarrow{u}_Q q'_Q$, where $q'_Q \in \Im(q'_\xi)$.

In all of the above mentioned cases, the second condition of bisimulation relation for the converse case holds true. Hence, $T_\xi$ and $T_Q$ are bisimilar. ∎

## 7. Adopting the DES supervisory control to the abstracted model

Following the above procedure, the partitioned system $T_Q$ was bisimilarly abstracted to a finite state machine $T_\xi$. The advantage of this method is that the control synthesis of this finite state machine can be effectively handled within Discrete Event Systems (DES) supervisory control theory initiated by Ramadge and Wonham [22].

### 7.1. DES model of the plant

The finite state machine $T_\xi$ can be formally presented by an automaton $G = (X, \Sigma, \alpha, X_0, X_m)$, where $X = X_\xi$ is the set of states; $X_0 = X_{\xi_0} \subseteq X$ is the set of initial states; $X_m = \{\tilde{R}_{1,j,k} | 1 \leq j \leq n_\theta - 1, \ 1 \leq k \leq n_\phi - 1\}$ is the set of final (marked) states; $\Sigma$ is the (finite) set of events. The sequence of these events forms a string. We use $\varepsilon$ to denote an empty string, while $\Sigma^*$ is as the set of all possible strings over the set $\Sigma$ including $\varepsilon$. The language of the automaton $G$, denoted by $L(G)$, is the set of all strings that can be generated by $G$ starting from the initial states. The marked language, $L_m(G)$, is the set of strings that belong to $L(G)$ and end with the marked states. $L(G(x_0))$ is the set of strings that belong to $L(G)$ and start from the initial state $x_0$. $\bar{L}$ is the set of all prefixes to the strings that belong to the language $L$.

Here, the event set $\Sigma$ consists of the actuation events $U_a = \{C_q^s | q \in \{r, \theta, \phi\}, s \in \{+, -\}\} \cup \{C_0\}$, the crossing the detection events $U_c = \{\hat{d}([i, j, k], [i', j', k']) | 1 \leq i, i' \leq n_r - 1, \ 1 \leq j, j' \leq n_\theta - 1, \ 1 \leq k, k' \leq n_\phi - 1\}$, and the external events $U_e$. Also, the set $U_e$ only contains the event $ca$ for the collision-alarm; Indeed, the event $ca$, will be generated when the system detects that another agent is located on the way of the follower towards the desired position. The event set $\Sigma$ consists of the controllable event set $\Sigma_c = U_a$ and uncontrollable event set $\Sigma_{uc} = U_d = U_c \cup U_e$. The uncontrollable events are those that cannot be affected by the supervisor. In automaton $G$, $\alpha : X \times \Sigma \to X$ is the transition function, which is a partial function and determines the possible transitions in the system caused by an event. This function is corresponding to $\to_\xi$ in $T_\xi$, so that for any $r \xrightarrow{v}_\xi r'$ we have $\alpha(r, v) = r'$. Based on the definition of $T_\xi$ and constructed controllers $C_0, C_r^+, C_r^-, C_\theta^+, C_\theta^+, C_\phi^+, C_\phi^-$, we have:

$$\alpha(\tilde{R}_{i,j,k},\sigma) = \begin{cases} \tilde{R}_{i,j,k} & \sigma = C_0 \\ \tilde{R}_{i,j,k} & \sigma = ca \quad for \ i \neq 1 \\ \tilde{d}([i,j,k],[i+1,j,k]) & \sigma = C_r^+ \quad for \ i \neq n_r - 1 \\ \tilde{d}([i,j,k],[i-1,j,k]) & \sigma = C_r^- \quad for \ i \neq 1 \\ \tilde{d}([i,j,k],[i,j+1,k]) & \sigma = C_\theta^+ \quad for \ j \neq n_\theta - 1 \\ \tilde{d}([i,j,k],[i,1,k]) & \sigma = C_\theta^+ \quad for \ j = n_\theta - 1 \\ \tilde{d}([i,j,k],[i,j-1,k]) & \sigma = C_\theta^- \quad for \ j \neq 1 \\ \tilde{d}([i,j,k],[i,n_\theta-1,k]) & \sigma = C_\theta^- \quad for \ j = 1 \\ \tilde{d}([i,j,k],[i,j,k+1]) & \sigma = C_\phi^+ \quad for \ k = n_\phi - 1 \\ \tilde{d}([i,j,k],[i,j,k-1]) & \sigma = C_\phi^- \quad for \ k = 1 \end{cases}$$

$$\alpha(\tilde{d}([i,j,k],[i',j',k']),\sigma) = \tilde{R}_{i',j',k'} \quad \sigma = \hat{d}([i,j,k],[i',j',k'])$$

Some parts of the graph representation of the system automaton are shown in Fig. 7. In this automaton, the arrows starting from one state and ending to another state represent the transitions, labeled by the events belong to $\Sigma$. The entering arrows stand for the initial states. As it is shown in Fig. 7, the system could start from any of the states $\tilde{R}_{i,j,k}$.

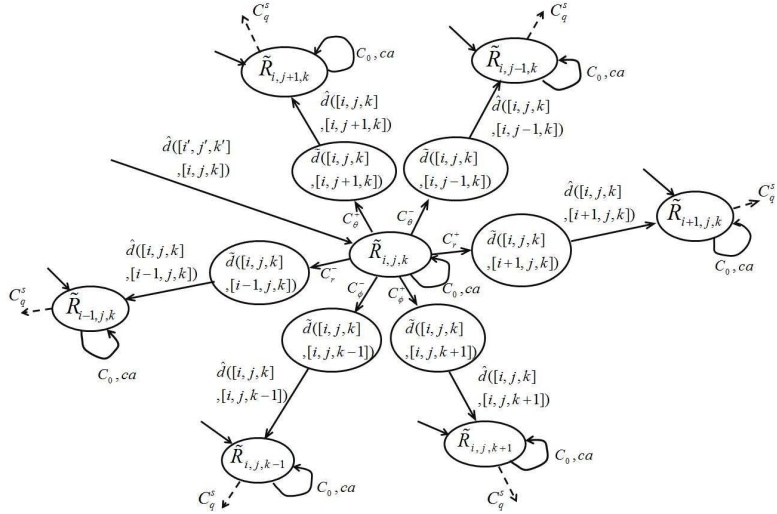

Figure 7: DES model of the plant.

In this model, when the system is in the state $\tilde{R}_{1,j,k}$, the command $C_r^-$ leads the system to an unknown region and the system would become non-deterministic. Hence, as it is reflected in the expression of $\alpha$, the command $C_r^-$ cannot be activated in this region. Similarly, the commands $C_\phi^+$ and $C_\phi^-$ cannot be activated in $\tilde{R}_{i,j,n_\phi-1}$ and $\tilde{R}_{i,j,1}$, respectively. Moreover, in the region $\tilde{R}_{n_r-1,j,k}$ the command $C_r^+$ cannot be activated as it leads the system outwards the control horizon. These restrictions are already considered in the definition of $\alpha$.

Now, the aim is to design a controller to bring the trajectories to the final states $R_{1,j,k}$, regardless of the initial state of the system. The control design issues are discussed in the next section. Before that, to simplify the DES model of the plant and to reduce its number of states, we can merge the states that have a similar situation in terms of the control labels that can be activated in these states. In this case, the refined model (Fig. 8) and the original DES model (Fig. 7) are language equivalent, meaning that $L(G) = L(G_{ref})$ and $L_m(G) = L_m(G_{ref})$ [32]. To do the refinement, first, all of the detection states $\{\tilde{d}([i,j,k],[i',j',k'])|\, 1 \leq i, i' \leq n_r - 1,\ 1 \leq j, j' \leq n_\theta - 1,\ 1 \leq k, k' \leq n_\phi - 1\}$ can be merged into a state called $D$. Then, one can merge similar states based on the events that can be activated for each region. Indeed, as described in the definition of $\alpha$, for the states that are adjacent to the sphere's boundary, the event $C_r^+$ cannot be activated. For the regions that touch the positive side of $z$ axis, negative side of $z$ axis, and the origin of the sphere, the events $C_\phi^-$, $C_\phi^+$, and $C_r^-$ cannot be activated, respectively. Following these rules, the states that are adjacent to the boundary of the control horizon and does not touch the $z$ axis, $\{\tilde{R}_{n_r-1,j,k}|1 \leq j \leq n_\theta - 1,\ 1 < k < n_\phi - 1\}$, can be aggregated to the state $N$ shown in Fig. 8. Other merged states are: $N_1 = \{\tilde{R}_{n_r-1,j,1}|1 \leq j \leq n_\theta - 1\}$, $N_n = \{\tilde{R}_{n_r-1,j,n_\phi-1}|1 \leq j \leq n_\theta - 1\}$, $R = \{\tilde{R}_{1,j,k}|1 \leq j \leq n_\theta - 1,\ 1 < k < n_\phi - 1\}$, $R_1 = \{\tilde{R}_{1,j,1}|1 \leq j \leq n_\theta - 1\}$, $R_n = \{\tilde{R}_{1,j,n_\phi-1}|1 \leq j \leq n_\theta - 1\}$, $P = \{\tilde{R}_{i,j,k}|1 < i < n_r - 1,\ 1 \leq j \leq n_\theta - 1,\ 1 < k < n_\phi - 1\}$, $P_1 = \{\tilde{R}_{i,j,1}|1 < i < n_r - 1,\ 1 \leq j \leq n_\theta - 1\}$, and $P_n = \{\tilde{R}_{i,j,n_\phi-1}|1 < i < n_r - 1,\ 1 \leq j \leq n_\theta - 1\}$. The list of the events that can be activated in each merged state, and fire a transition to the detection state $D$ form the set $\gamma_c(*)$, which is defined as follows:

$$\gamma_c(r) = \begin{cases} \{C_r^+, C_r^-, C_\theta^+, C_\theta^-, C_\phi^+, C_\phi^-\} & r = P \\ \{C_r^+, C_r^-, C_\theta^+, C_\theta^-, C_\phi^+\} & r = P_1 \\ \{C_r^+, C_r^-, C_\theta^+, C_\theta^-, C_\phi^-\} & r = P_n \\ \{C_r^-, C_\theta^+, C_\theta^-, C_\phi^+, C_\phi^-\} & r = N \\ \{C_r^-, C_\theta^+, C_\theta^-, C_\phi^+\} & r = N_1 \\ \{C_r^-, C_\theta^+, C_\theta^-, C_\phi^-\} & r = N_n \\ \{C_r^+, C_\theta^+, C_\theta^-, C_\phi^+, C_\phi^-\} & r = R \\ \{C_r^+, C_\theta^+, C_\theta^-, C_\phi^+\} & r = R_1 \\ \{C_r^+, C_\theta^+, C_\theta^-, C_\phi^-\} & r = R_n \end{cases} \qquad (15)$$

Transition from the detection state $D$ to one of the merged states generates the detection events that are presented by the set $\gamma_d(*)$. For instance $\gamma_d(P) = \{\hat{d}([i,j,k],[i',j',k'])|\, 1 \leq i \leq n_r - 1,\ 1 \leq j, j' \leq n_\theta - 1,\ 1 \leq k \leq n_\phi - 1,\ 1 < i' < n_r - 1,\ 1 < k' < n_\phi - 1\}$ and $\gamma_d(P_1) = \{\hat{d}([i,j,k],[i',j',1])|\, 1 \leq i \leq n_r - 1,\ 1 \leq j, j' \leq n_\theta - 1,\ 1 \leq k \leq n_\phi - 1,\ 1 < i' < n_r - 1\}$.

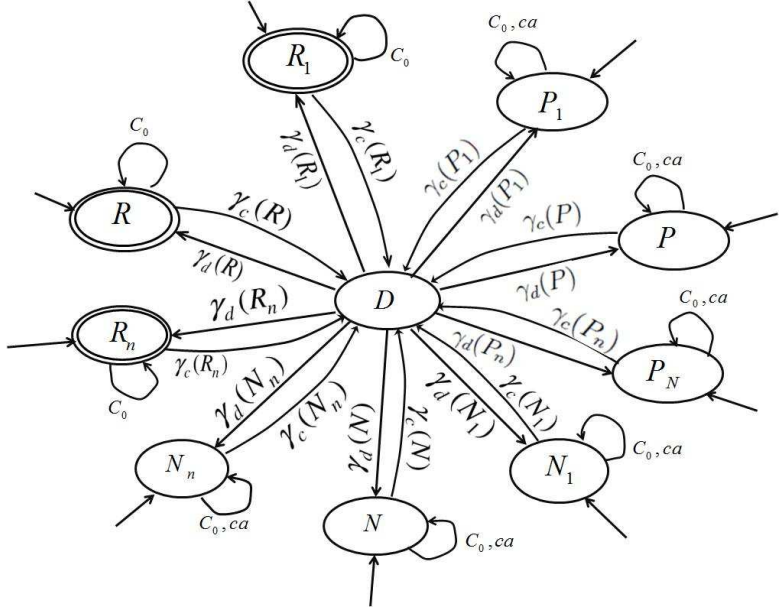

Figure 8: Refined DES model of the plant.

### 7.2. Design of the supervisor

The logical behavior of the system can be modified by a discrete supervisor to achieve a desired order of events. Indeed, the supervisor, $S$, observes the executed events of the plant $G$ and disables the undesirable controllable strings. Here, we assume that all of the events are observable. The language and marked language of the closed-loop system, $L(S/G)$ and $L_m(S/G)$, can be constructed as follows:

(1) $\varepsilon \in L(S/G)$
(2) $[(s \in L(S/G)) \text{ and } (s\sigma \in L(G)) \text{ and } (\sigma \in L(S))] \Leftrightarrow (s\sigma \in L(S/G))$
(3) $L_m(S/G) = L(S/G) \cap L_m(G)$

where $s$ is the string that has been generated so far, and $\sigma$ is an event, which the supervisor should decide whether keep it active or not.

Within this framework we can use parallel composition to facilitate the control synthesis. Parallel composition is a binary operation between two automata. Here, the parallel composition is used to combine the plant discrete model and the supervisor.

**Definition 5.** (**Parallel Composition**)[33] Given $G = (X_G, \Sigma_G, \alpha_G, x_{0_G}, X_{m_G})$ and $S = (X_S, \Sigma_S, \alpha_S, x_{0_S}, X_{m_S})$, $G_{cl} = G\|S = (X_{cl}, \Sigma_{cl}, \alpha_{cl}, x_{0_{cl}}, X_{m_{cl}})$ is said to be the parallel composition of $G$ and $S$ with $X_{cl} = X_G \times X_S$, $\Sigma_{cl} = \Sigma_G \cup \Sigma_S$, $x_{0_{cl}} = (x_{0_G}, x_{0_S})$, $X_{m_{cl}} = X_{m_G} \times X_{m_S}$, and $\forall x = (x_1, x_2) \in X_{cl}, \sigma \in \Sigma_{cl}$, then $\alpha_{cl}(x, \sigma) =$

$$\begin{cases} \bullet(\alpha_G(x_1, \sigma), \alpha_S(x_2, \sigma)) \\ \quad if\ \alpha_G(x_1, \sigma)!\ and\ \alpha_S(x_2, \sigma)!\ and\ \sigma \in \Sigma_G \cap \Sigma_S \\ \bullet(\alpha_G(x_1, \sigma), x_2) \quad if\ \alpha_G(x_1, \sigma)!\ and\ \sigma \in \Sigma_G - \Sigma_S \\ \bullet(x_1, \alpha_S(x_2, \sigma)) \quad if\ \alpha_S(x_2, \sigma)!\ and\ \sigma \in \Sigma_S - \Sigma_G \\ \bullet undefined \quad otherwise \end{cases}$$

where $\alpha_*(x, \sigma)!$ shows the existence of a transition from the state $x$ by the event $\sigma$ in system $*$. In this definition, the initial condition of the automata was assumed to be singular. Extending this definition for the case that automata $G$ and $S$ have the initial sets $X_{0_G}$ and $X_{0_S}$, the initial set of the composed system will be $X_{0_{cl}} = \Upsilon(X_{0_G}, X_{0_S}) \subseteq X_{0_G} \times X_{0_S}$, where the relation $\Upsilon$ describes the initial states in $G$ and $S$ that are coupled to synchronously generate a string in the composed system.

In fact, parallel composition synchronizes operand systems on their comment events, however, their private events can transit independently. Next lemma and Corollary use the parallel composition to obtain the closed-loop system.

**Lemma 3.** *[33] Let $G = (X, \Sigma, \alpha, x_0, X_m)$, be the plant automaton with the initial state of $x_0$ and $K \subseteq \Sigma^*$ be a desired language. There exists a nonblocking supervisor $S$ such that $L(S/G) = L(S||G) = K$ if and only if $\emptyset \neq K = \bar{K} \subseteq L(G)$ and $K$ is controllable. In this case, $S$ could be any automaton with $L(S) = L_m(S) = K$.*

Using the above lemma and following the definition of the parallel composition, the result can be extended to a plant with several initial states:

**Corollary 1.** *Let $G = (X, \Sigma, \alpha, X_0, X_m)$ be the plant with the initial state set $X_0 = \{x_0^1, x_0^2, ...\}$ and $K = \bigcup K_i \subseteq \Sigma^*$ be a desired language, where $K_i$ is the desired language that should be generated starting from $x_0^i$. If $\emptyset \neq K_i = \bar{K_i} \subseteq L(G(x_0^i))$ and $K_i$ is controllable for all $i = 1, 2, ..., |X_0|$, there exists a nonblocking supervisor $S$ such that $L(S/G) = L(S||G) = K$. In this case, $S$ could be any automaton that has the initial state set $S_0 = \{s_0^1, s_0^1, ..., s_0^m\}$, $m \leq |x_0|$, and for any $x_0^i$ there exists a $s_0^j$, $(x_0^i, s_0^j) \in \Upsilon$, which satisfies $L_m(S(s_0^j)) = L(S(s_0^j)) = K_i$, where $\Upsilon$ is the coupling relation between the initial states of supervisor $S$ and the plant $G$.*

Now, using the parallel composition and the above corollary, we will design the supervisor for reaching the formation, keeping the formation, and collision avoidance modularly.

*7.2.1. Design of the controller for reaching and keeping the formation*

For reaching the formation, it is sufficient to drive the follower UAV directly towards one of the regions $R_{1,j,k}$, $1 \leq j \leq n_\theta - 1$, $1 \leq k \leq n_\phi - 1$, located in the first sphere. After reaching $R_{1,j,k}$, the UAV should remain inside it, forever. This specification, $K_F$, is realized in Fig. 9. In this figure, the initial state $P_f$ is considered to be coupled with the states $P$, $P1$, $P_n$, $N$, $N_1$, $N_n$, in Fig. 8. Being in one of these states, the UAV is not in the first sphere, and the event $C_r^-$ will be generated to push the UAV towards the origin. Entering into a new state, the event $d([i, j, k], [i', j', k'])$ will appear to show the current state of the system. This will continue until the event $d([i, j, k], [1, j', k'])$ be generated, which shows that the formation is reached. In this case, the event $C_0$ will be activated, which keeps the system trajectory in the first region. Since, there is another module to handle the collision avoidance, the formation supervisor does not change the generatable language after the event $ca$, and lets the collision avoidance supervisor disable undesirable events as it will be explained in the next section. It can be seen that $K_F$ is controllable as it does not disable any uncontrollable event.

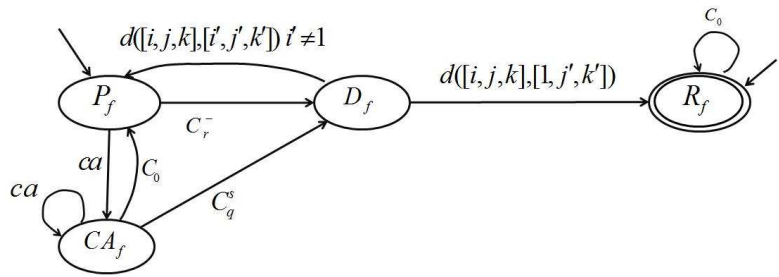

Figure 9: The realization of the reaching and keeping the formation specification.

Based on Corollary 1, there exists a supervisor that can control the plant to achieve this specification. The supervisor is the realization of the above specification in which all states are marked. Marking all states of the supervisor allows the closed-loop marked states be solely determined by the plant marked states. The supervisor for reaching the formation and keeping the formation is denoted by $S_F$. The closed-loop system can be obtained using the parallel composition: $G_{cl} = S_F/G = S_F || G$. Here, the coupling relation is as follows: $\Upsilon = \{(R_1, R_f), (R, R_f), (R_n, R_f), (P_1, P_f), (P, P_f), (P_n, P_f), (N_1, P_f), (N, P_f), (N_n, P_f)\}$. All of the events are common between the plant and the supervisor. Moreover, it can be seen that $L(S_F) \subseteq L(G_{ref})$. Therefore, knowing that $L(G) = L(G_{ref})$, it will be concluded that:

$$\begin{aligned} L(S_F/G) &= L(G||S_F) = L(G) \cap L(S_F) = \\ &\quad L(G_{ref}) \cap L(S_F) = L(S_F) = K_F \end{aligned} \qquad (16)$$

**Remark 3.** *If for any reason the follower UAV deviates from the required position and for instance, enters into one of the adjacent regions nondeterministically, the event $d([i, j, k], [i', j', k'])$ will be generated, which informs that the UAV is entered into the region $R_{i',j',k'}$. Having this information,* *the controller can resume its job from the newly entered region and push the UAV towards the origin.*

7.2.2. Collision Avoidance Supervisor

When the follower UAV is going to reach the desired position, in some situations the follower may collide with the leader. More precisely, when the follower UAV is in the region $R_{i,j,k}$ and the leader UAV is located in the region $R_{i',j,k}$, and $i' < i$, then the collision alarm, $ca$, will be generated. If we look at this problem from the relative frame point of view, the leader UAV has a fixed position in this frame. Therefore, when the follower detects that the leader is located on its path towards the desired position (center of sphere), it suffices that the follower turns to change its azimuth angle, $\theta$, by activating the event $C_\theta^+$. When, it enters into the region $R_{i,j+1,k}$, the event $d([i,j,k],[i,j+1,k]) \in U_d$ will appear, and the collision alarm will be removed. In this case, the collision avoidance supervisor lets the formation supervisor resume the reaching the formation. To do so, the collision avoidance supervisor only changes the generatable language after happening the event $ca$ and lets the rest be treated by the formation supervisor. The collision avoidance supervisor, $S_C$, is shown in Fig. 10. Here, the coupling relation is $\Upsilon = \{(P_1, S_C), (P, S_C), (P_n, S_C), (N_1, S_C), (N, S_C), (N_n, S_C)\}$. Again, all the events are common between the plant and the supervisor. Hence, $L(S_C) \subseteq L(G_{ref})$ and $L(G) = L(G_{ref})$ lead to:

$$\begin{aligned} L(S_C/G) &= L(G||S_C) = L(G) \cap L(S_C) \\ &= L(G_{ref}) \cap L(S_C) = L(S_C) = K_C \end{aligned} \quad (17)$$

where $K_C$ is the collision avoidance specification.

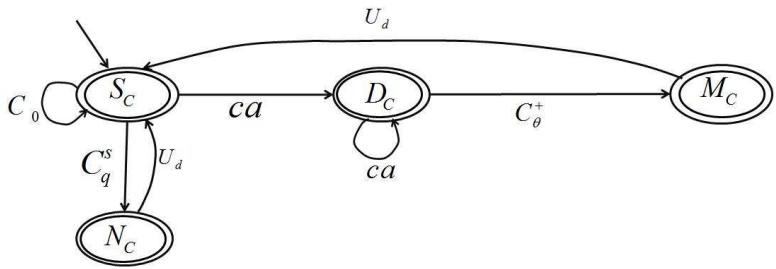

Figure 10: Collision avoidance supervisor, $S_C$.

*7.2.3. The closed-loop system*

For prefix closed languages $K_F$ and $K_C$, we can apply modular synthesis [33], using composition of the plant, the reaching and keeping the formation, and the collision avoidance supervisor:

$$G_{cl} = G||S_F||S_C \tag{18}$$

Here, the closed-loop language can be achieved as follows:

$$\begin{aligned}L(G||S_F||S_C) &= L(G) \cap L(S_F) \cap L(S_C) = \\ L(S_F) &\cap L(S_C) = K_F \cap K_C\end{aligned} \tag{19}$$

The refined closed-loop automaton, $G_{cl_{ref}}$, is shown in Fig. 11.

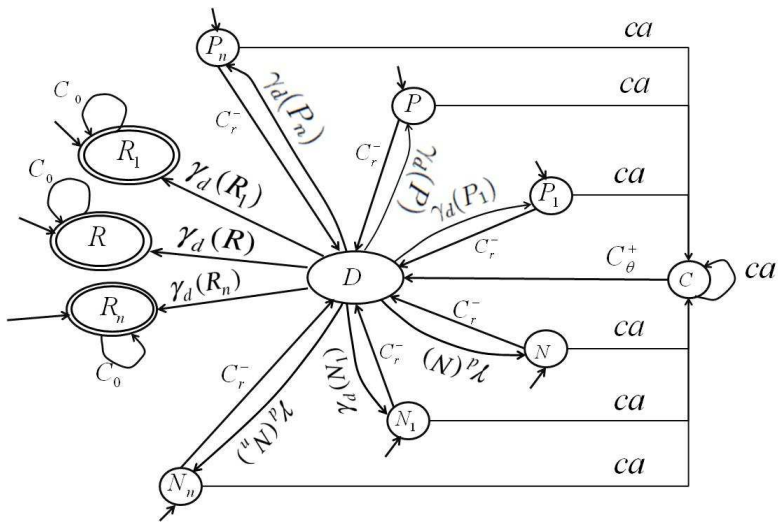

Figure 11: The closed-loop system.

## 8. Construction of the hybrid controller

To describe the partitioned system and its relation with the discrete supervisor, analogous with [19] and referring to the definition of $T_Q$, we can define an interface layer (Fig. 12(a)), which connects the supervisor to the plant. The interface layer has two main blocks: the *detector* and the *actuator*. The detector converts the continuous time signals to a sequence of symbols. Upon crossing the detection elements, a plant symbol, $\hat{d}([i,j,k],[i',j',k'])$, is generated, which informs the supervisor about the current situation of the plant. Based on the observed plant symbols, the supervisor decides, which control signal should be given to the plant to satisfy the desired specification. This command has a discrete nature, but the control commands to be given to the plant need to be continuous. The actuator translates the discrete commands to the continuous ones:

$$r(v) = u(x) = \Sigma_m \lambda_m(x) u(v_m) \qquad (20)$$

where $v$ is the discrete command and $u(v_m)$ should be chosen from the sets $U_m(Inv(R_{i,j,k}))$ if $v = C_0$. Otherwise it should be selected from the set $U_m(Ex(F_q^s(R_{i,j,k})))$ for $v = C_q^s$. The coefficients $\lambda_m(x)$, $m = 0, ..., 7$, can be obtained from (5). The whole structure, the interface layer and the supervisor, is implemented on the follower UAV.

Indeed, the plant and the interface layer elements, together with the actuator and the detector, form the transition system $T_Q$. It was shown that this transition system can be bisimilarly abstracted to the finite state machine $T_\xi$ (Fig. 12(b)) for which we designed the discrete supervisor. Due to the bisimilarity of $T_Q$ and $T_\xi$, the designed supervisor for $T_\xi$ can work for $T_Q$ so that the closed-loop system behavior does not change, leading to:

**Theorem 5.** *With the aid of interface layer, the discrete supervisor $S = S_C || S_F$ can be applied to the original partitioned system, $T_Q$, so that the closed-loop system satisfies the required specification, $K_F \cap K_C$.*

## 9. Hardware-in-the-loop simulation results for the formation algorithm

To verify the proposed algorithm we have used a hardware-in-the-loop simulation platform [34] developed for NUS UAV helicopters [27]. In this platform, the nonlinear dynamics of the UAV has been replaced with its nonlinear model of the UAV, and all software and hardware components that are involved in a real flight test, remain active during the simulation. In this platform, the control commands and references will be sent to the actuators, but the helicopter does not move as the motor is off. Indeed, the variables of the nonlinear model will change, and instead of the measured values, these values will be given back to the controller to compute new control values. Consequently, the simulation results of this simulator is very close to the actual flight tests. The controller is embedded in a $PC \backslash 104$ and the controller details and program code is explained in [35].

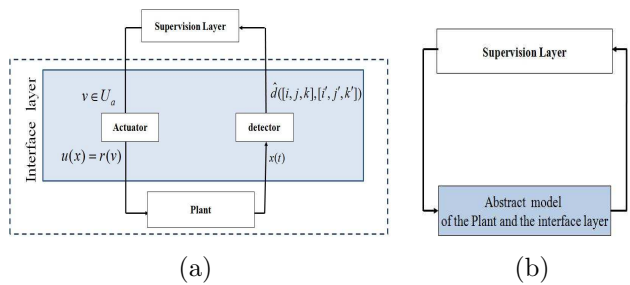

Figure 12: (a) Interface layer structure and (b) Abstract model for the plant and interface layer.

Now, consider two UAVs each of which are simulated by the hardware-in-the-loop software. The leader follows a predefined path and the follower receives the leader's data via a wireless communication board (IM-500X008, FreeWave). This multi-UAV simulator test bed is used to simulate the algorithm in two cases:

1. Scenario 1: To monitor reaching the formation behavior of the UAV, we have used a fixed leader and the follower should reach the desired position.
2. Scenario 2: To monitor how the follower is able to maintain the achieved formation, the leader tracks a circle path, and the follower should reach and keep the formation.

9.1. Simulation of reaching the formation and collision avoidance

The system dynamics, $\dot{x} = u$, is a multi-affine function. The controller, $u(x)$, drives the UAV inside a spherical partitioned space. The control signal is generated using the control mechanism described in Section 5. The control horizon is a sphere of diameter 50m. The partitioning parameters are selected as $n_r = 15$ and $n_\theta = 20$ and $n_\phi = 10$. To construct the controllers $C_0$, $C_r^+$, $C_r^-$, $C_\theta^+$, $C_\theta^-$, $C_\phi^+$, and $C_\phi^-$, we can apply Theorem 2 and 3, respectively. For example, to construct the controllers $C_0$ and $C_r^-$, it is sufficient to choose the values of $u(v_m)$ from the sets $U_m(Inv(R_{i,j,k}))$ and $U_m(Ex(F_r^-(R_{i,j,k})))$ given in (13) and (14), respectively.

To simulate reaching the formation stage, assume that the leader has a fixed position and the follower should reach a desired position that has a certain distance from the leader. Also assume that $R_m < d$ so that collision would not be happen. Considering that the follower has a relative distance $(dx, dy, dz) = (17, 18, 8)$ with respect to the desired position, reaching the formation stage behavior is shown in Figure 13. The projection of the UAV position onto the $x-y$ plane is shown in Fig. 14(a). The follower UAV's state variables are shown in Fig. 15(a). Moreover, the regions that the follower UAV has crossed are presented in Fig. 14(b), where $i$, $j$, $k$ are the indices of the traversed regions $R_{i,j,k}$. As it can be seen, the UAV has finally reached the final state $R_{1,13,5}$. The control signals are shown in Fig. 15(b).

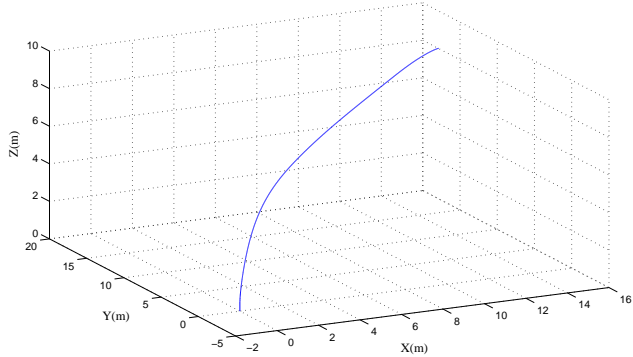

Figure 13: Reaching the formation stage.

Now, with the similar situation as above, assume that the relative distance between the follower and the desired position is $(dx, dy, dz) = (-17, -18, -8)$. With the partitioning parameters $R_m = 50m$, $n_r = 15$, $n_\theta = 20$, and $n_\phi = 10$, the initial state of the system is $R_{8,13,5}$. Assume that the leader is located in $R_{7,13,5}$. Since the leader is located on the path of follower towards the desired position, a collision avoidance alarm will be generated to activate the collision avoidance mechanism. The collision avoidance behavior of the system is shown in Fig. 13. The projection of the UAV position onto the $x - y$ plane is shown in Fig. 16(b). The state variables of the system are shown in Fig. 17(b). The traversed regions are depicted in Fig. 17(a), where $i$, $j$, $k$ are the indices of the crossed regions. The follower, first has moved towards the region $R_{8,14,5}$ to avoid the collision, then it has resumed reaching

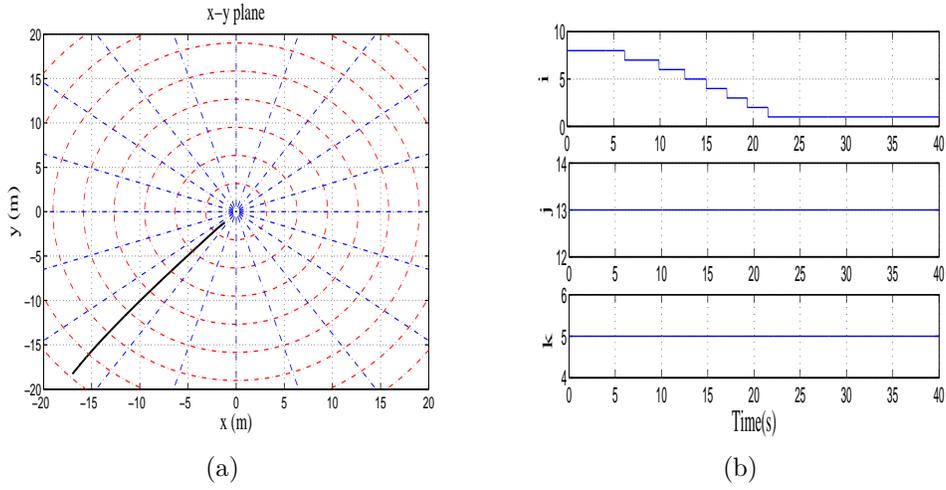

Figure 14: (a) Reaching the formation stage projected onto the x-y plane and (b) The traversed regions' indices in reaching the formation stage.

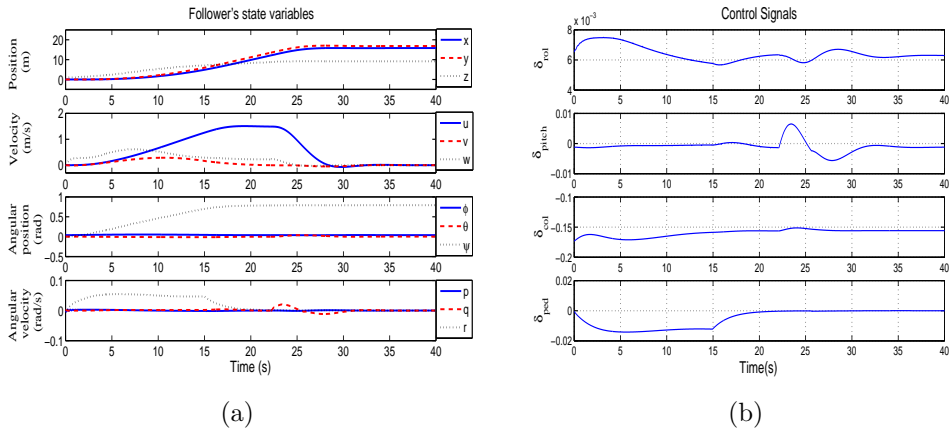

Figure 15: (a) The state variables of the follower UAV in reaching the formation stage and (b) Control input signals in reaching the formation stage.

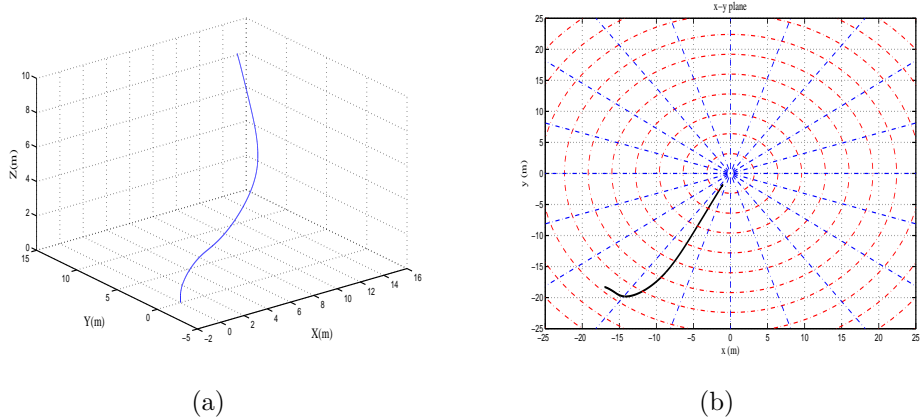

(a) (b)

Figure 16: (a) The position of the follower UAV for the collision avoidance mechanism, (b) Relative distance of the follower from the desired position, projected onto the x-y plane.

the formation to complete the mission.

9.2. Simulation of keeping the formation

To monitor reaching and keeping the formation, let the leader to track a circle with the diameter of 20m and with the altitude of 20m. Consider the partitioning parameters as the above scenario. Here, the follower is initially located at $(dx, dy, dz) = (-20, -20, -20)$ with respect to the leader. It is expected that after a while, the follower reaches the relative distance of $(dx, dy, dz) = (5, 5, 5)$ with respect to the leader. The behavior of the follower UAV is shown in Fig. 18(a). The follower UAV's state variables are shown in Fig. 18(b). As it can be seen the follower has finally reached the desired formation and has successfully kept it.

9.3. Simulation of a multi-follower scenario

The algorithm can be extended to a multi-follower case. Each of the followers has his own supervisor, which makes it able to reach the formation, keep the formation, and avoid the collision. Each follower just needs to have the position and velocity information of the leader and also the position of the neighbor agents. In this case, the reaching the formation and keeping the formation is similar to what we designed so far, but a more complicated collision avoidance is required. We will consider the multi-follower case in our future research works and here, we just show how the idea can be used for a simple multi-follower case.

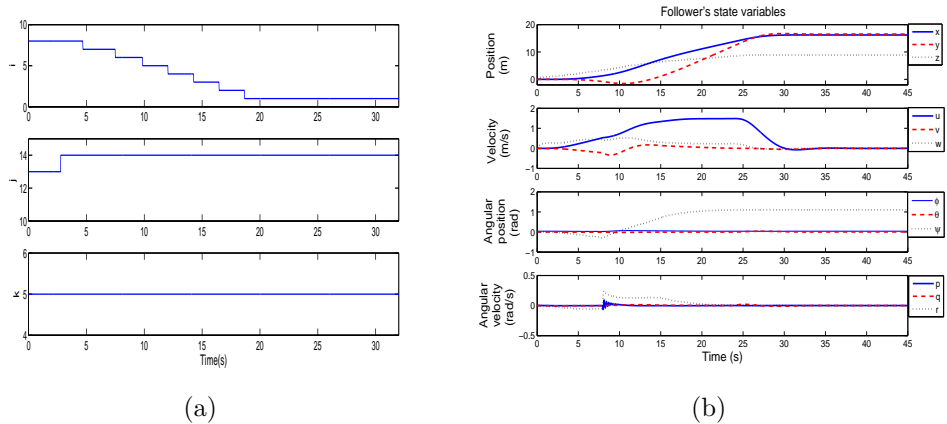

Figure 17: (a) The traversed regions' indices for the collision avoidance mechanism and (b) The state variables of the follower UAV during the collision avoidance mechanism.

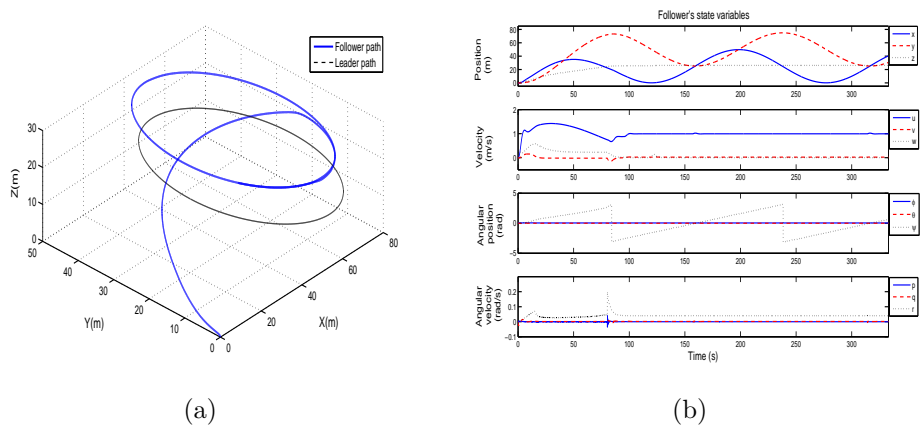

Figure 18: (a) The position of the UAVs in a circle formation mission and (b) The follower UAV's state variables during a circle formation mission.

Here, the partitioning parameters are $R_m = 100m$, $n_r = 20$, $n_\theta = 5$. Followers 1 is initially located at a point, which has a relative distance of $(dx, dy, dz) = (-30.1, 12.1, -12.3)$ with respect to the leader, while the desired relative distance is $(dx, dy, dz) = (-15, -5, 10)$. Follower 2 is initially located at a point which has a relative distance of $(dx, dy, dz) = (3.8, 33.4, 0)$ with respect to the leader. The desired relative distance between this UAV and the leader is $(dx, dy, dx) = (-15, -5, 10)$. The position of the UAVs in the 3-dimensional space is shown in Fig. 19 and the projection of their position onto the x-y plane is shown in Fig. 20. As it can be seen, each UAV has reached the formation and has kept it successfully.

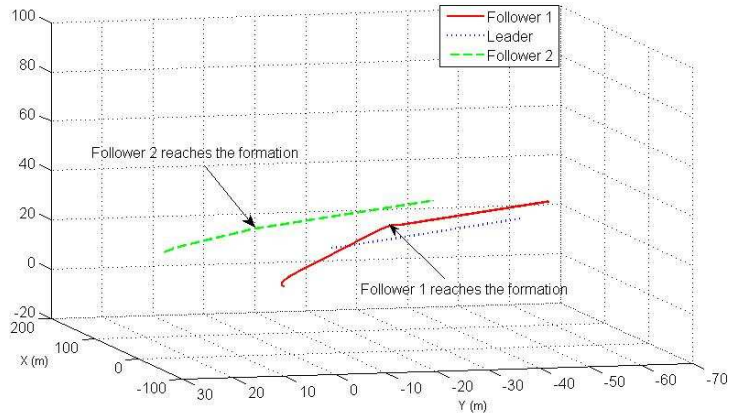

Figure 19: Multi-follower formation control.

## 10. Conclusion

In this paper, a hybrid supervisory control was proposed for the three-dimensional leader-follower formation control of unmanned helicopters. The approach was based on spherical abstraction of the state space and utilizing the properties of multi-affine functions over the partitioned space. The finite state DES model of the plant was achieved by the proposed abstraction procedure, and then, a discrete supervisor was modularly designed to satisfy the desired specifications. The designed supervisor is able to form the formation, maintain the achieved formation, and take care of the inter-collision between the agents. To link the discrete supervisor and the continuous model of the plant, an interface layer was embedded. Due to the bisimulation relation between the partitioned plant and its DES model, it is guaranteed that the DES model and the original hybrid model of the plant exhibit the same behavior. The algorithm was verified by a hardware-in-loop simulator and the presented results demonstrate the effectiveness of the designed controller.

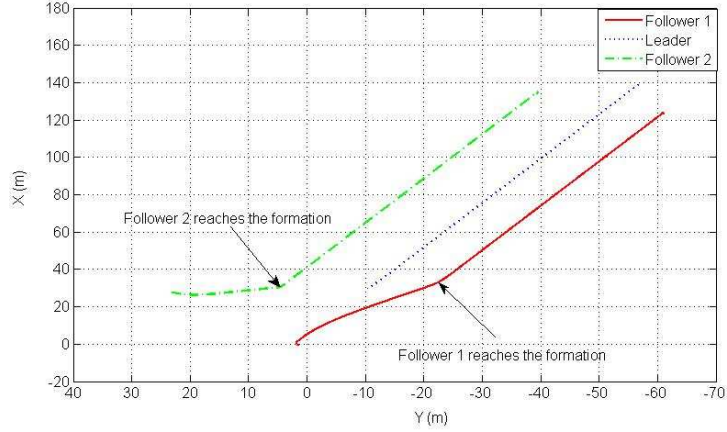

Figure 20: The projection of the UAVs' position onto the x-y plane in multi-follower scenario.

## 11. APPENDIX

*11.1. Proof for Theorem 1*

let $x = (r, \theta, \phi) \in \bar{R}_{i,j,k}$. Therefore, from the partitioning procedure: $r_i \leq r \leq r_{i+1}$, $\theta_j \leq \theta \leq \theta_{j+1}$, and $\phi_k \leq \phi \leq \phi_{k+1}$. Hence, $r$, $\theta$, and $\phi$ could be written affinely as:

$$\begin{cases} r = (1-\lambda_r)r_i + \lambda_r r_{i+1} & 0 \leq \lambda_r \leq 1 \Rightarrow \lambda_r = \frac{r-r_i}{r_{i+1}-r_i} \\ \theta = (1-\lambda_\theta)\theta_j + \lambda_\theta \theta_{j+1} & 0 \leq \lambda_\theta \leq 1 \Rightarrow \lambda_\theta = \frac{\theta-\theta_j}{\theta_{j+1}-\theta_j} \\ \phi = (1-\lambda_\phi)\phi_k + \lambda_\phi \phi_{k+1} & 0 \leq \lambda_\phi \leq 1 \Rightarrow \lambda_\phi = \frac{\phi-\phi_k}{\phi_{k+1}-\phi_k} \end{cases}$$

Consider a trajectory from the vertex $v_0 = v_{000}$ towards the point $x$ only moving along the spherical axis. As an example, $v_0 = (r_i, \theta_j, \phi_k) \xrightarrow{step1} x_1 = (r, \theta_j, \phi_k) \xrightarrow{step2} x_2 = (r, \theta, \phi_k) \xrightarrow{step3} x_3 = x = (r, \theta, \phi)$. In fact, in each step, we have changed only one parameter, and fixed the others to take the advantages of the multi-affine functions. When in the function $g(r, \theta, \phi)$, the parameter $\theta$ and $\phi$ are fixed and only $r$ is varying, we use the notation $g_{\theta\phi}$ to highlight the fixedness of $\theta$ and $\phi$. Similarly, we can define $g_{r\theta}$ and $g_{r\phi}$. As $g$ is a multi-affine function, $g_{\theta\phi}$, $g_{r\theta}$, and $g_{r\phi}$ are affine. Since in step1, the parameters $\theta$ and $\phi$ are fixed and only $r$ is changing:
$g(x_1) = g_{\theta\phi}|_{\theta_j,\phi_k}(r) = g_{\theta\phi}|_{\theta_j,\phi_k}((1-\lambda_r)r_i + \lambda_r r_{i+1}) = (1-\lambda_r)g_{\theta\phi}|_{\theta_j,\phi_k}(r_i) + \lambda_r g_{\theta\phi}|_{\theta_j,\phi_k}(r_{i+1}) = (1-\lambda_r)g(r_i,\theta_j,\phi_k) + \lambda_r g(r_{i+1},\theta_j,\phi_k)$.

In step 2, the parameter $\theta$ changes and the other parameters are fixed. Therefore:
$g(x_2) = g_{r\phi}(\theta)|_{r,\phi_k} = g_{r\phi}|_{r,\phi_k}((1-\lambda_\theta)\theta_j + \lambda_\theta \theta_{j+1}) = (1-\lambda_\theta)g_{r\phi}|_{r,\phi_k}(\theta_j) + \lambda_\theta g_{r\phi}|_{r,\phi_k}(\theta_{j+1}) = (1-\lambda_\theta)g(r,\theta_j,\phi_k) + \lambda_\theta g(r,\theta_{j+1},\phi_k) = (1-\lambda_\theta)g_{\theta\phi}|_{\theta_j,\phi_k}(r) + \lambda_\theta g_{\theta\phi}|_{\theta_{j+1},\phi_k}(r)$.

In step one, we have already obtained $g_{\theta\phi}|_{\theta_j,\phi_k}(r)$. With the same procedure, we can obtain $g_{\theta\phi}|_{\theta_{j+1},\phi_k}(r)$. Applying these two values in step 2, $g(x_2)$ will be:
$g(x_2) = (1-\lambda_\theta)[(1-\lambda_r)g(r_i,\theta_j,\phi_k) + \lambda_r g(r_{i+1},\theta_j,\phi_k)] + \lambda_\theta[(1-\lambda_r)g(r_i,\theta_{j+1},\phi_k) + \lambda_r g(r_{i+1},\theta_{j+1},\phi_k)] = (1-\lambda_\theta)(1-\lambda_r)g(r_i,\theta_j,\phi_k) + (1-\lambda_\theta)\lambda_r g(r_{i+1},\theta_j,\phi_k) + \lambda_\theta(1-\lambda_r)g(r_i,\theta_{j+1},\phi_k) + \lambda_\theta\lambda_r g(r_{i+1},\theta_{j+1},\phi_k)$

In Step 3, the parameter $\phi$ changes and the other parameters are fixed. Therefore: $g(x_3) = g_{r\theta}(\phi)|_{r,\theta} = (1-\lambda_\phi)g(r,\theta,\phi_k) + \lambda_\phi g(r,\theta,\phi_{k+1})$

where $g(r,\theta,\phi_k)$ is obtained in Step 2 and $g(r,\theta,\phi_{k+1})$ can be obtained in a similar way. Substituting these values in $g(x_3)$ and rearranging the coefficients, the result will be as (4) and (5).

*11.2. Proof for Proposition 2*

The existence already guaranteed by Theorem 1. For the proof of uniqueness, assume by contradiction that $f$ is not unique, and there is another multi-affine function $f'$ that $f(v_m) = f'(v_m) = g(v_m)$, or equivalently, $f''(v_m) = f(v_m) - f'(v_m) = 0$. It could be easily proven, that if $f$ and $f'$ are multi-affine, $f'' = f - f'$ also is multi-affine. By Theorem 1, $\forall x \in R_{i,j,k} : f''(x) = \sum \lambda_m f''(v_m) = 0$, $m = 0,2,...,7$. Therefore, $\forall x \in R_{i,j,k}, f(x) = f'(x)$.